\documentclass{aa}
\usepackage{graphicx}
\usepackage{multirow}
\usepackage{txfonts}
\usepackage{floatflt}
\usepackage[authoryear]{natbib}
\bibpunct{(}{)}{;}{a}{}{,}
\usepackage{amstext}

\begin{document}

\title{Model computations of blue stragglers and W UMa-type stars in
  globular clusters}
\author{K. St\c epie\'n
\and M. Kiraga}

\institute{Warsaw University Observatory, Al.~Ujazdowskie~4, 
    00-478~Warsaw, Poland, email: kst, kiraga@astrouw.edu.pl}
\date{Received; accepted}
\abstract
{It was recently demonstrated that contact binaries occur in globular
  clusters (GCs) only immediately below turn-off point and in the 
  region of blue straggler stars (BSs). In addition, observations 
  indicate that at least a
  significant fraction of BSs in these clusters was formed
  by the binary mass-transfer mechanism.} 
{The aim of our present investigation is to obtain and analyze
a set of
evolutionary models of cool, close detached binaries with a
low metal
abundance, which are characteristic of GC.}
 {We computed the evolution of
975 models of initially detached, cool
  close binaries 
with different initial parameters. The models include mass exchange
between components as well as mass and angular momentum
loss due to the magnetized winds for very low-metallicity binaries
with Z = 0.001. The models are interpreted in the context of
existing
data on contact binary and blue straggler members of GCs. The models are based on our recently developed code for
evolutionary modeling of cool close and contact binaries.}
 {The model parameters agree well with the observed positions {{ of
     the GC contact binaries}}
   in the Hertzsprung-Russell
   diagram, and we achieve detailed parameters of {{ several}} individual contact 
   and near-contact binaries. Contact binaries in the lower part of the cluster main sequence 
   are {{ absent}} because  there are {{ no}} binaries with initial orbital periods shorter than
   1.5 d. Contact
   binaries end their evolution as mergers that appear in the  BS
   region. {{ Binary-formed}} BSs populate the whole observed BS
   region in a GC, {{ but a gap is visible between low-mass mergers
     that are concentrated along the zero-age main sequence and binary BSs occupying the red part
     of the BS region.}}
   Very few binary mergers are  expected to rotate rapidly and/or possess
   chemical peculiarities resulting from the exposure of the layers
   processed by CNO nuclear reactions. All other binary mergers are
   indistinguishable from the collisionally formed mergers. The
   results show that
   {{ binary-formed BSs may constitute}} at least a substantial fraction of
   all BSs in a GC.} 
{}
\keywords{binaries: close - blue stragglers - stars:
  late type - stars: evolution - globular clusters: general} 

\titlerunning{Blue stragglers and W UMa-type stars in globular clusters}
\authorrunning{K. St\c epie\'n and M. Kiraga}
\maketitle

\section{Introduction}

W UMa-type stars are solar-type eclipsing variables with nearly
sinusoidal light curves in which both binary components share a common
envelope that is located between the inner and outer critical surfaces {
  (equipotential zero-velocity surfaces crossing the L1 and L2 Lagrangian
  points)}, also called inner and outer Roche lobes
\citep{moch81}. The orbital periods of W UM-type stars are 
  concentrated around
0.25-0.6~d.  Because of the physical properties of W UMa-types stars, we also call them contact binaries (CB).

Numerous W UMa-type stars are observed in globular and old open
clusters \citep{ruc98,ruc00}, but they are completely absent among T
Tauri stars and in young clusters,{ even though they are easy to
  detect \citep{math94,melo01,heb10}.  None is known in the
  intermediate-age cluster Hyades, in spite of the high degree of
  completeness achieved in the binary search by \citet{grif85},} and
only one certain W UMa-type member is identified in Praesepe
\citep{ruc98}. These data indicate that it takes several Gyr -- up to
the age of a globular cluster (GC) -- before a CB is formed from its
progenitor. The age of several Gyr also results from the kinematic
analysis of field W UMa-type stars \citep{gbra88, bil05}.

It is generally accepted that progenitors of CBs are cool detached
binaries. In the course of evolution they lose angular momentum (AM)
by a magnetized wind blowing from one or both components. 
With synchronous rotation, stellar AM loss (AML) results in a
tightening of the orbit until the primary fills its inner Roche lobe and Roche-lobe overflow (RLOF) occurs. For a binary whose initial
orbital period is close to 2 d and whose initial primary is similar to the
Sun, the time to reach RLOF is approximately equal to the
evolutionary life time of the primary on the main sequence (MS) \citep{ste11a}. For shorter
initial periods and/or lower primary mass, the MS life time is
longer. Nonetheless, as long as the initial
orbital period is not very different from 2 d, the solar type initial primary
is already advanced in evolution when RLOF occurs.

W UMa-type stars observed in GCs always lie close to the turn-off (TO)
in the Hertzsprung-Russell diagram (HRD), with a significant fraction
falling above TO, in the region of blue straggler
stars (BSs), which are stars lying on the high-mass extension
of the cluster MS \citep{ruc00}. Early observations of GCs were not
sufficiently sensitive to detect CBs in the lower part of the cluster
MS, but the recent deep surveys of a few GCs revealed an absence of CBs
fainter than about one magnitude below TO \citep{kath09, kaetal10,
  kaetal13a}. As \citet{kaetal13a} stressed, not a single system was
detected in over 12 000 stars in the M4 cluster with magnitudes 17.8 $<
V <$ 21, that is, between one and four magnitudes below TO,
whose light curves were analyzed. 

The aim of our present investigation is to obtain and analyze a set of
evolutionary models of cool, close, detached binaries with a low metal
abundance, which {\bf is} characteristic of GC. Depending on the specific initial
conditions, a fraction of the investigated binaries changes into CBs
at the later age. Evolutionary computations are based on a model
developed by one of us \citep{ste95,ste06a, gste08, ste09}. Altogether,
about $10^3$ models have been computed with an initial primary mass
between 0.7-0.9 $M_{\sun}$, an initial secondary mass between 0.4-0.8
$M_{\sun}$, an initial orbital period between 1.5-2.9 d, and a metal
abundance $Z$ = 0.001. The results are compared with the available
observations of GCs, in particular with the data on the cluster CBs
and BSs.

The next section presents the essentials of the evolutionary model, including the recent improvements of the computer
program, which follow the evolution of both components more accurately
than in the original code (in particular, the slow mass-transfer rate
is now calculated every time step), and additional
effects are included. The third section describes the models
together with the statistical properties of the cool close binary
population as it evolves from the beginning to an{ age of 13.5 Gyr}. 
Bulk properties of binaries that form mergers or are in a contact
state at an age of 11, 12, and 13 Gyr are discussed. The observed properties
of CBs and BSs, including individual objects from different GCs, are
compared with the model predictions in the fourth section.  The last
section summarizes and discusses the results of the paper.

\section{Description of an evolutionary model of cool 
close binaries} 

\subsection{Equations}

The cool close binary model { (CCBM)} describes the evolution of a cool close
binary from the ZAMS untill a stage immediately preceding the
merger of the
components or the formation of an Algol-type star with an orbital period
of several days. We do not follow the detailed evolution of the Algol-type binary here because it does not fulfill some of the CCBM
assumptions. In addition, the Algol and post-Algol evolution has been
investigated thoroughly in the past and is well known
\citep[e.\thinspace g.,][]{pacz71,sar93,negg01}.

As an initial close binary we took a detached system with an orbital
period short enough to ensure full synchronization of the orbital and
rotational periods. Because we are mostly interested in progenitors of
CBs, we discuss only binaries with  initial periods shorter
than 3 d. Longer period models always evolve into Algols. Cool
means here that both components possess subphotospheric convection
zones that generate magnetic activity, in particular, a magnetized stellar
wind. An approximate upper mass limit for these stars is about 1.3
$M_{\sun}$, although somewhat more massive stars can also be
considered because they later develop a convection zone during their MS
life. We here focus on stars
with initial masses { equal to or} lower than 0.9 $M_{\sun}$. They have a
sufficiently deep convection zone already on the ZAMS. We apply the Roche
model to describe the orbital parameters.

The basic equations of CCBM are the third Kepler
law, the expression for AM, and the approximate expressions for inner
Roche-lobe sizes $r_1$ and $r_2$ \citet{eggl83},

\begin{equation}
P = 0.1159a^{3/2}M^{-1/2}\,,
\label{kepler}
\end{equation}

\begin{equation}
H_{\rm{tot}} = H_{\rm{spin}} + H_{\rm{orb}}\,,
\label{totam}
\end{equation}

where 

\begin{equation}
H_{\rm{spin}} = 7.13\times10^{50}(k^2_1M_1R^2_1 +k^2_2M_2R^2_2)P^{-1}\,,
\label{spin}
\end{equation}

and

\begin{equation}
H_{\rm{orb}} = 1.24\times 10^{52}M^{5/3}P^{1/3}q(1+q)^{-2}\,,
\label{orbam}
\end{equation}

\begin{equation}
\frac{r_1}{a} = \frac{0.49q^{2/3}}{0.6q^{2/3}+\ln(1 + q^{1/3})}\,,
\end{equation}

\begin{equation}
\frac{r_2}{a} = \frac{0.49q^{-2/3}}{0.6q^{-2/3}+\ln(1 +
q^{-1/3})}\,.
\end{equation}

Here $P$ is the period (rotational and orbital), $M = M_1 + M_2$ is
the total mass with component masses $M_1$ and $M_2$, $R_1\text{and
} R_2$ are component radii, $a$ is the semi-axis,
$H_{\rm{tot}}$, $H_{\rm{spin}}$ and $H_{\rm{orb}}$ are total,
rotational and orbital AM, $k_1^2$ and $k_2^2$ are
the (nondimensional) gyration radii of the two components, $r_1
 \text{and
} r_2$ are the
sizes of the inner Roche lobes, and $q = M_1/M_2$ is the mass
ratio. Masses, radii, and semi-axis are given in solar units, the period is given in days, and the AM
in cgs units.

We assumed that stellar winds from the two components are the
dominating mechanism of the orbit evolution. They carry
away mass and AM according to the formulas

\begin{equation}
\dot M_{1,2} = -10^{-11}R_{1,2}^2\,,
\label{massloss}
\end{equation}

\begin{equation}
\frac{{\rm d}H_{\rm{tot}}}{{\rm d}t} = -4.9\times 10^{41}
(R_1^2M_1 +R_2^2M_2)/P\,.
\label{amloss}
\end{equation}

Here $\dot M$ is in solar mass per year and 
${{\rm d}H_{\rm{tot}}}/{{\rm d}t}$ is in gcm$^2$s$^{-1}$ per year. The
formulas were discussed at length by \citet{ste06a,ste06b} and
\citet{gste08}. They are calibrated by the observational data of the
rotation of single, magnetically active stars and empirically
determined mass-loss rates of single, solar type stars. Both formulas
apply in a limiting case of a rapidly rotating star in the saturated
regime. Note that they do not contain any free adjustable
parameters. The constant in Eq.~(\ref{massloss}) is uncertain within a
factor of 2 and that in Eq.~(\ref{amloss}) is uncertain to $\pm$
30\% \citep{ste06b,wood02}. Any interaction between winds from the two components is neglected.

\subsection{Initial assumptions}

The model equations were applied to binaries{ with parameters
  characteristic of GC members. The CCBM evolution was followed until the
  age of the oldest GCs and snapshots of the model population
  properties are discussed in detail for three specified ages of 11, 12
  and 13 Gyr.} We focused { on} stars with initial masses
close to the TO mass, because more massive cluster members have
already completed their evolution and have formed compact objects, whereas the
less massive ones are still on the MS, burning hydrogen in their
cores. Specifically, we took initial primary masses from the interval
0.7-0.9 $M_{\sun}$, with the detailed values given in
Table~\ref{init}. If the primary in a model has a mass that is closely spaced
between 0.86-0.87 $M_{\sun}$ {  , it means that the primaries
  reach RLOF  between 11 and 13 Gyr. The initial masses of the
secondary component
   were adjusted to avoid extreme mass ratios $q_{\mathrm{init}}
  = 1$ and $q_{\mathrm{init}}>> 1$. The evolution of binaries with exactly
  equal mass components is uninteresting to us, whereas the rapid mass
  transfer following RLOF is very likely nonconservative for high
  values of the mass ratio. Without a reliable theory describing
  this process, the standard procedure is to use free parameters for
  mass and AM losses \citep{sar96}. To avoid introducing them, we
  assumed conservative mass transfer. Our models have $1.08
  \le q_{\mathrm {init}} \le 2.25$}. The masses of the secondary component are also given in Table~\ref{init}. For each pair of the
component masses, 15 models were computed with initial period
lengths from 1.5 d to 2.9 d, every 0.1 d. Altogether, 975 evolutionary
models were obtained. In the following, we { identify} individual
models by giving the initial values of three basic parameters:
$M_1+M_2(P)$, for instance, 0.8+0.6(1.8).

\begin{table}
\caption{Initial component masses of the computed binary models. For
  each combination of $M_1$ and $M_2$ a set of 15 evolutionary models has
been calculated with initial orbital periods 1.5, 1.6,...,2.8,2.9 d}
\label{init}
\centering 
\begin{tabular}{ll}
\hline
\hline
 $M_1 (M_{\sun})$ & $M_2 (M_{\sun})$ \\
\hline
 0.7 & 0.4, 0.5, 0.6 \\  
 0.75 & 0.4, 0.5, 0.6 \\
 0.8 & 0.4, 0.5, 0.6, 0.7 \\
 0.81 & 0.4, 0.5, 0.6, 0.7 \\      
 0.82 & 0.4, 0.5, 0.6, 0.7 \\
0.83 & 0.4, 0.5, 0.6, 0.7 \\
0.84 & 0.4, 0.5, 0.6, 0.7 \\
0.85 & 0.4, 0.5, 0.6, 0.7 \\
0.86 & 0.4, 0.5, 0.6, 0.7 \\
0.865 & 0.4, 0.5, 0.6, 0.7, 0.8 \\
0.8675 & 0.4, 0.5, 0.6, 0.7, 0.8 \\
0.87 & 0.4, 0.5, 0.6, 0.7, 0.8 \\
0.88 & 0.4, 0.5, 0.6, 0.7, 0.8 \\
0.89 & 0.4, 0.5, 0.6, 0.7, 0.8 \\
0.9 & 0.4, 0.5, 0.6, 0.7, 0.8 \\
\hline
\end{tabular} 
\end{table}

We adopted the range of initial period values according to the
following considerations.  It is
now generally accepted that binary and multiple stars are formed during
the early fragmentation of a protostellar cloud
\citep{boss93,bonn94,krat10,mach08}. After the end of the rapid accretion
phase, cool protostars enter the T Tauri phase when the mass has
almost reached its final value and changes
only little. This transition takes place at the age of $10^5-10^6$
years \citep{mach10,vor10,oht13}. The orbit of a binary with two
freshly formed T Tauri-type components must be wide enough to
accommodate both stars. The youngest T Tauri stars with masses between
0.5 and 1.5 $M_{\sun}$ have radii from 2 up to over 3 solar radii
\citep{bar98,tog11}, which means that the initial orbital period must be longer than
about 1.5 d (the accurate value depends only weakly on the mass
ratio). We adopted this value as a lower limit for the initial orbital
period of an isolated binary. Shorter period binaries can be formed by
dynamical interactions (collisions) with other objects in a GC
\citep{heg75,hil75,hut83,lei13, hyp13}, but we excluded them from our analysis.

There exists another efficient mechanism that effectively shortens
the orbital periods of some binaries. It requires a third
body on a very wide and strongly inclined orbit around the inner
binary. The third companion induces so-called Kozai cycles that
periodically change the eccentricity of the inner orbit. Tidal interactions
between the components during periastron passages dissipate the system
energy and tighten the orbit. Numerical models of this mechanism,
called Kozai cycles with tidal friction (KCTF), show that a
significant fraction of binaries with initial periods of 2-3 weeks
shorten their periods to 2-3 d within several million years
\citep{eggl06,fab07,per09,nao14}. At these periods the orbits
circularize and do not change any more \citep{fab07}. Population
synthesis indicates that among binaries with periods shorter than 3 d,
those formed by the KCTF mechanism outnumber isolated systems
\citep{fab07}. Observations of field binaries confirm this
expectation. \citet{tok06} found out that 96\% of binaries with
periods shorter than 3 d possess a third companion, in contrast to
34\% of binaries with periods longer than 12 d. A similar conclusion
was reached by \citet{rpk07}, who showed that all or almost all field W
UMa-type binaries possess a tertiary component.

Consequently, we assumed that the initial period
distribution of cool close binaries has a sharp maximum at 2-3 d,
resulting mostly from the KCTF mechanism, and a rapid decline toward 1.5
d. The extension of the distribution beyond 3 d is unimportant for
our purposes because, as we shown below, these binaries
do not form CBs.

\subsection{Description of evolutionary computations}

The evolution of cool close binaries can be divided into three phases
\citep{ste06a,pacz07,gste08}, where phase I begins at the initial state
to RLOF by the initial primary (denoted with the subscript ``1''
throughout, also when it becomes a {\em \textup{less}} massive
component). The other component is marked ``2''. The binary is
detached in phase I, and its evolution is dominated by two mechanisms:
individual evolution of each component across the MS, and the magnetized
winds carrying away mass and AM. Any interaction between winds from
the two components is neglected, and no other mechanism influencing the
orbit is considered. Equations~(\ref{kepler}-\ref{amloss}) are integrated
at every time step. The typical length of the time step was set to $6.8\times
10^7$ years so that the age of a GC is reached in about 200 steps. At each
time step, the radii of the components were calculated using the grid of
single-star models with a metal content $Z$ = 0.001 published by
\citet{gir00}. The value of the radius was interpolated in time and
mass. The total mass lost during phase I amounts to several percent of
the initial stellar mass, depending on{ stellar radius and} the
duration of phase I. This has a minor effect on the radius evolution
of the components. More important is the evolutionary advancement due to
hydrogen burning, which slowly inflates a star. At the same time,
the Roche lobes descend onto the stellar surfaces due to AML
(Eq.~\ref{amloss}). If RLOF did not occur within 13.5 Gyr,
calculations were stopped and the binary remained detached. To allow for
the supersaturation of magnetic activity in ultra-fast rotators
\citep{ran96,pro96,ste01}, the period in Eq.~(\ref{amloss})
was replaced by 0.4 d for binaries with orbital periods shorter than
that limiting value. Phase I is the longest evolutionary phase with
a duration of the same order as the MS lifetime of the initial primary.

Following RLOF, phase II begins when a rapid mass transfer occurs from
the initial primary (henceforth loser) to the initial secondary
(gainer). It is much shorter than other phases and lasts no more
than one or a few percent of phase I. The details of this process are
discussed by \citet{stekir13a}. The gainer swells after receiving a
small amount of matter, until it fills its inner Roche lobe and the
system temporarily assumes a contact configuration \citep{webb76}. The
mass that then flows from the loser encircles the swollen gainer close to
its equator and returns to the parent star \citep{ste09}. With no
mass accreted, the swollen star slowly shrinks, trying to revert to
thermal equilibrium.  When its surface moves below the Roche lobe, the
star can again receive a small amount of mass, which causes it to expand
back to the size of the lobe. The process of shrinking, gaining some
mass, and expanding again is repeated (a sort of equilibrium
is established where an instantaneous mass-accretion rate just
balances swelling and shrinking of the star close to the Roche lobe),
resulting in a mass-transfer rate governed by the thermal timescale
of the gainer rather than loser. As soon as the thermal timescale of
the loser becomes longer than that of gainer (this occurs when the two
components have approximately equal masses), the gainer can receive all the
transferred mass.{ Phase II ends when both stars regain
  equilibrium and the loser just fills its Roche lobe. The mass ratio is
  inverted so that the initial primary is now less massive than the initial
  secondary. Regardless of the details of the mass transfer process, we
  assumed that mass is transferred on a thermal timescale of $\sim 10^8$
  years. A constant mass transfer rate $\dot M_{\rm{tr}} = 5\times
  10^{-9} M_{\sun}$ yr$^{-1}$ was adopted based on observation that
  the total mass transferred during phase II in any of the modeled
  binary did not exceed 0.5~$M_{\sun}$.

  In most cases, a short-period semi-detached configuration of the
  Algol-type was obtained at the end of phase II with the gainer
  inside its Roche lobe. We call this a near contact binary (NCB). But a
  few binaries were so compact that a CB was already formed before
  phase III began.

  Similarly as in phase I, the evolutionary expansion of both stars,
  accompanied by mass and AM loss by the winds, drives the binary
  evolution in phase III. In particular, the expansion of the loser beyond
  its Roche lobe results in a slow mass transfer (on the evolutionary
  timescale) to the gainer. This widens the orbit, which in
  turn suppresses the transfer. In absence of winds, the binary
  period systematically increases until mass transfer stops
  completely. In the presence of a low AML rate, the period still
  increases, although at a lower pace. If the AML rate is high  and/or the mass transfer rate low, however, the orbit shrinks until both
  components merge. Each binary follows one of these routes,
  depending on the relative efficiency of the two processes.}

\begin{figure}
\includegraphics[height=\hsize]{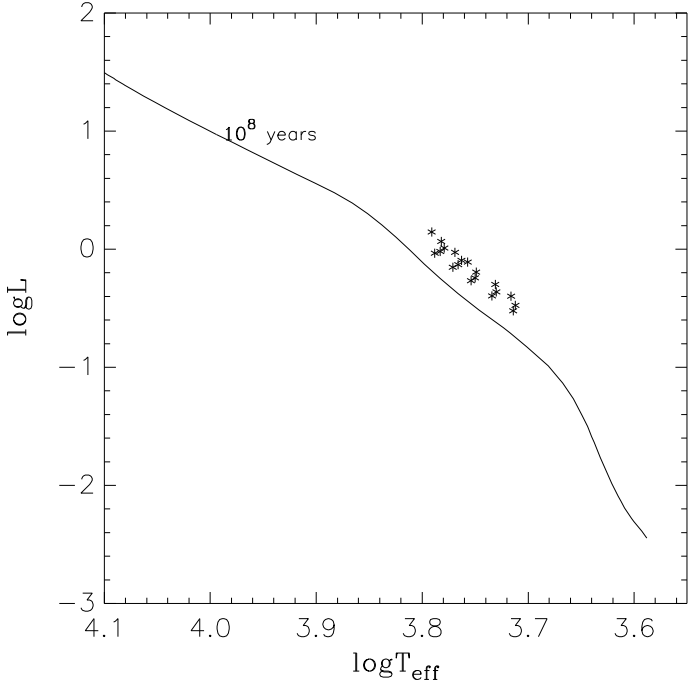}
\caption{HR diagram with representative initial binary models plotted
  together with a low-metal isochrone assumed to track the ZAMS of a GC.}
\label{inhr}
\end{figure}

 To compute the binary evolution in phase III, we used the same
 equations and assumptions as in phase I. Particularly important was a
 proper treatment of the loser. Because we lack
 detailed evolutionary models of binary components after a rapid mass exchange,
 single-star models were used, as described above. Analytic formulas
 were fitted to mass-radius relations of stars at different
 evolutionary stages: at the MS, at the TAMS, and for several
 values of the helium core mass. In addition, formulas for time derivatives of
 stellar radii were found for different masses and evolutionary
 stages. Using them, the radius of the secondary component was
 computed at each time step and compared to the Roche lobe size.{ The
 mass transfer rate resulted from this comparison.

 Several different formulas for $\dot M_{\rm{tr}}$ following RLOF in
 cool close binaries have been used in the past assuming conservative
 mass transfer. \citet{ibtu84}
 assumed proportionality of $\dot M_{\rm{tr}}$ to $(R_1/r_1)^{10}$,
 \citet{eggl02} to ${(\ln (R_1/r_1))^3}$, whereas \citet{han00} used}


\begin{eqnarray}
\dot M_{\rm{tr}}  = & C\frac{(R_1 - r_1)^{\alpha}}{r_1} & {\rm for}\enskip R_1 > r_1\,,\\
\dot M_{\rm{tr}} = & 0  & {\rm for} \enskip R_1 \leq r_1\,,\nonumber
\label{masstransfer}
\end{eqnarray}

{ with $C$ = 1000 $M_{\sun}$yr$^{-1}$ and $\alpha$ = 3. After
  several trials, we adopted the formula of Han et al., but with $C$ = 1
  $M_{\sun}$yr$^{-1}$ and $\alpha$ = 5.} The adopted values secured
stability for most of the models. Only in a few cases unnecessary
oscillations appeared (see below).

This method of calculating the mass-transfer rate was { applied to
  semi-detached binaries} until the gainer reached its Roche lobe,
thus producing a CB. After that, another approach had to be used because
both stars overfill their Roche lobes. Now, the degree of overfill by
the gainer was calculated at every time step and compared with the
same quantity for the loser. If the { former} was lower than the
{ latter}, the mass transfer rate was increased { according to
  Eq.~9, except that an upper limit of $2\times 10^{-9}$
  $M_{\sun}$yr$^{-1}$ was imposed to keep the changes of the binary
  configuration smooth. If not, the zero transfer rate was applied for
  the current time step}. The model calculations were stopped when the
degree of the inner Roche-lobe overfill by both stars{ reached 20\%,
  attaining thus the outer Roche lobe for typical values of the
mass ratio
  of the modeled binaries ($q \approx 0.6-0.8$).} At this moment, the
binary was assumed to lose mass and AM through the Lagrangian L$_2$ point
at a rapidly increasing rate with an ultimate merging of the
components.  Similarly, merging should also occur as a result
of the Darwin
instability when the spin AM exceeds one third of the orbital AM. This occurs
for mass ratios below 0.1 \citep{rasio95}.  We stopped the
calculations{ at this value}, notwithstanding the low degree of
overfill.

The contact configuration{ arises from the NCB formed at the
  beginning of phase III} when the influence of AML on the orbit
prevails, or at least balances the mass-transfer rate so that the orbit
remains tight during phase III. In the opposite situation, the orbit widens,
the period increases, and the binary remains{  in the Algol configuration
  with the gainer} well inside its Roche lobe. This occurs when the
initial primary reaches RLOF after completing its MS evolution, so that
hydrogen in the central part of its core is already
exhausted. Consequently, it
moves to the red giant region on the HRD, and its rapid expansion
results in a high enough mass-transfer rate to dominate the orbit
evolution. We stopped the calculations for orbital periods exceeding 3
d when the spin-orbit synchronization may already break down.

The spin AM of the two components is about two orders of magnitude lower than
the orbital AM for all considered binaries during phases I-II and for
almost all binaries during phase III. Nonetheless, they were calculated
at each time step. Constant values of gyration radii were adopted:
$k_1^2 = 0.10,$ which corresponds to a star with core hydrogen depleted,
and $k_2^2 = 0.06$, corresponding to an unevolved MS star
\citep{rut88,clar92}. The value of $k_1^2$ is overestimated for early
evolutionary phases, but it better reflects the internal structure of a
loser in phase III. In addition, the total binary luminosity and the
average effective temperature of CCBMs were calculated at each time
step. { The} sum of the component luminosities was adopted as the binary
luminosity, and the binary effective temperature was obtained from the
surface-weighted mean brightness of both stars,

\begin{equation}
L_{\rm b} = L_1 + L_2\,,
\label{binlum}
\end{equation}

\begin{equation}
T_{\rm{eff,b}}^4 = L_{\rm b}/{4\pi\sigma (R_1^2 + R_2^2)}\,,
\label{bintemp}
\end{equation}

where $L_{\rm b}$ is the binary luminosity and $T_{\rm{eff,b}}$ is the
binary effective temperature.

Figure \ref{inhr} shows the region of the HRD that is occupied by the initial CCBMs. To
avoid crowding, only 19 models were plotted with a primary mass
between 0.7-0.9 $M_{\sun}$, every 0.1 $M_{\sun}$, and all secondary
masses accompanying them.
We also plot the isochrone $t = 10^8$ years, $Z = 0.001$ from the
latest Padova models ({\it http://stev.oapd.inaf.it/cmd}), which
shows the approximate position { of} the ZAMS for a GC.
The modeled binaries all lie above the cluster MS, as expected.

\section{Results}
\subsection{Period evolution}

\begin{figure}
\includegraphics[height=\hsize]{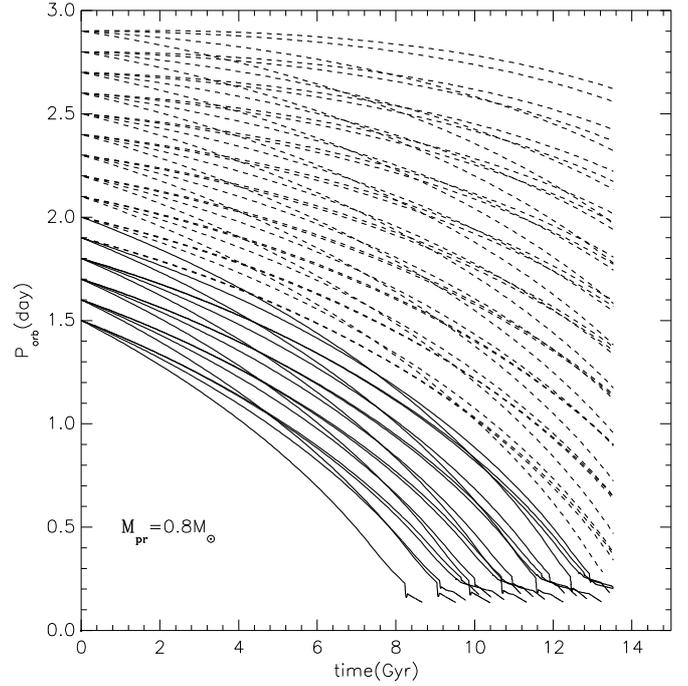}
\caption{Variation in time of the orbital period of all binary models
 with an initial primary mass equal to 0.8 $M_{\sun}$. Period
 variations of binaries that have not yet reached RLOF within the age of{ the Universe} are shown with dotted lines and solid lines describe the period 
variations of binaries that reach RLOF within this age.}
\label{P0.80}
\end{figure}

\begin{figure}
\includegraphics[height=\hsize]{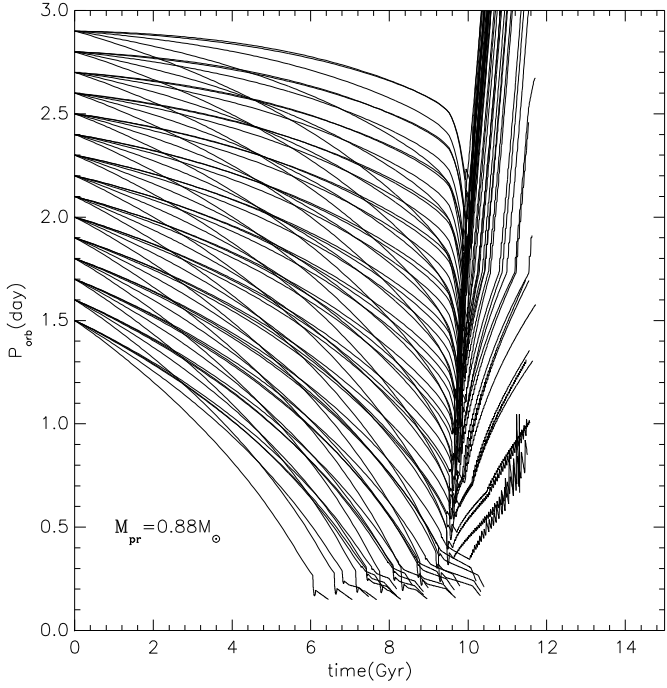}
\caption{Variation in time of the orbital period of all binary models
 with an initial primary mass equal to 0.88 $M_{\sun}$. All these
 binaries reached RLOF within the age of{ 11 Gyr}.}
\label{P0.88}
\end{figure}

\begin{figure}
\includegraphics[height=\hsize]{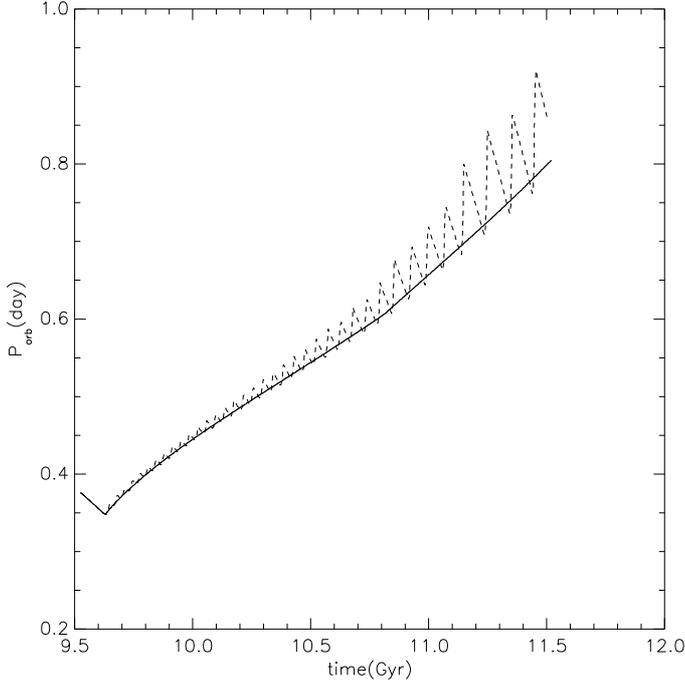}
\caption{Period variations of a model 0.88+0.4(2.2) in phase III,
calculated with a
  time step of $5\times 10^6$ years (dotted line) and $1.25\times
  10^6$ years, i.\thinspace e. four times shorter (solid line).}
\label{damp}
\end{figure}

The binary evolution sensitively depends sensitively on
whether the primary mass is higher or lower than the TO mass. Primaries
with lower masses remain detached untill the specified age of a GC, unless their
initial orbital period is{ significantly shorter than 1.5 d}.
Figure~\ref{P0.80} presents an example of the period evolution of binaries
with a{  primary mass lower than the TO mass. Out of all 60 models with an initial primary mass of 0.8 $M_{\sun}$
  , 19 (32\%) reach phase III within an age of 13.5
  Gyr,} whereas the others remain detached. Corresponding figures for
other low masses look similar, except that the number of models
reaching phase III increases from only 4 (9\% ) for 0.7 $M_{\sun}$ 
to 35 (58\%) for 0.86 $M_{\sun}$ (note that because of the mass lost through stellar winds, the final masses of the stars are about 6-7\%
lower than initial).  All binaries with initial primaries
more massive than 0.86 $M_{\sun}$ reach RLOF within the age of{ 13.5
  Gyr} and show pronounced evolutionary effects (see an example in
Fig.~\ref{P0.88}). Here we{ also} see that binaries with periods
longer than $\sim$ 2 d reach RLOF when their primaries have 
already left the
MS, which results in a high
mass-transfer rate in phase III and a rapid period increase. Some
models with $q$ approaching 0.1 show spurious period oscillations
before merging, see for example the model in the { lower right} of Fig.~\ref{P0.88}. This is caused by a too long time step. We
used a constant value of $5\times 10^6$ years for all models in phase III, which means that the typical duration of this phase is 200-400 steps. A high
instantaneous mass-transfer rate, multiplied by the long time step,
results in a rapid decrease of an already low{ loser} mass,
producing a { sudden period jump, followed by its decrease at
  the next time step as the mass transfer rate drops to zero.} 
We obtained exemplary
calculations for the time step shortened four times, that is,
for $1.25\times 10^6$ years. The period oscillations are
then completely absent (Fig.~\ref{damp}). Switching to such a short
time step would, however, substantially lengthen the computations,
therefore we decided to ignore these oscillations in the following
analysis. { We did so because a correct value of a binary period
  matters only when summation over consecutive bins is performed to
  obtain the period distribution at a specified age (see below). Instead
  of trying to determine a correct period value, for instance,
by determining the local
  minimum in the vicinity of this age, we included its
  instantaneous value in the summation. In effect, a small
  number of periods fall incorrectly into an adjacent bin. This changes
  the general appearance of the
  distribution only little. The components merge as a result
of the Darwin
  instability always near the local period minimum, which
  is close to the correct value (Fig.~\ref{damp}).}


\subsection{Period distribution}

The orbital period distribution of cool close binaries in GC evolves
in time under the influence of several mechanisms. Some, such
as encounters with other stars, harden the binaries by shortening their
periods \citep{heg75,hil75,hut83,hyp13,lei13}, or like the
KCTF mechanism that operates on a timescale of several Gyr because
of the very long initial period of the inner binary \citep{fab07}, inject
binaries into the considered period interval. Other mechanisms
such as expelling
binaries from GC, component merging, or formation of Algols
with longer periods, remove binaries from that interval. A full
discussion of all these mechanisms is beyond the scope of the present
paper. We consider here only the evolution of the period
distribution of CCBMs.


\begin{figure}
\resizebox{\hsize}{!}{\includegraphics{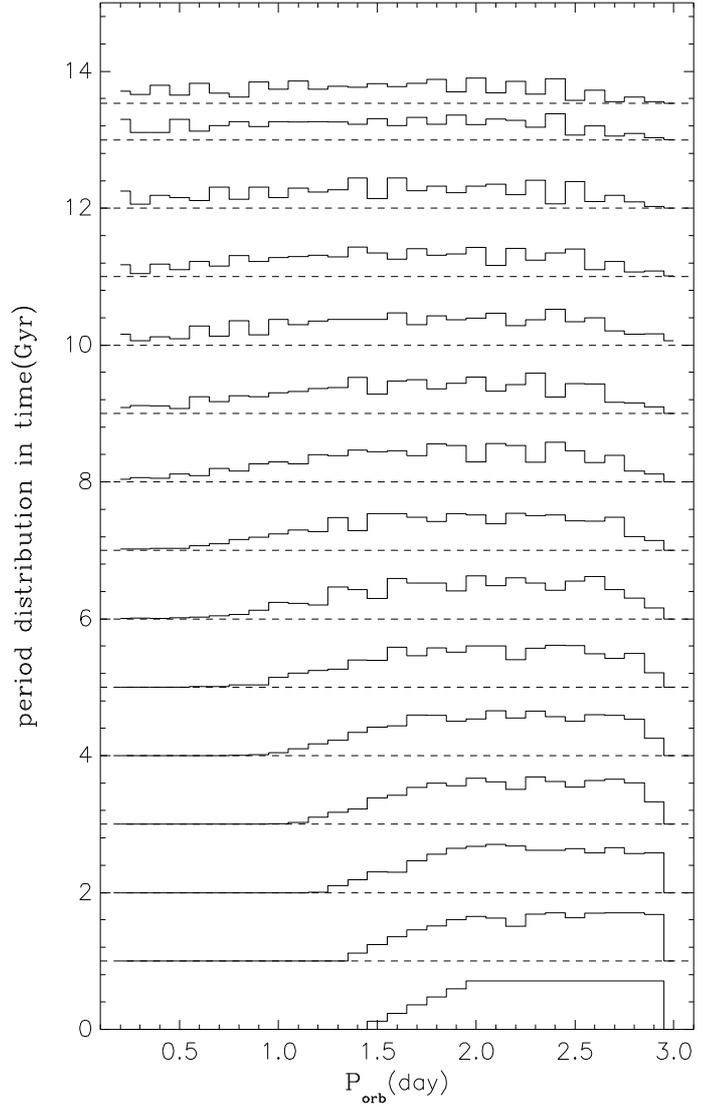}}
\caption{Evolution of the orbital period distribution of cool close binaries in
  time. Distributions are shown every 1 Gyr from 0 till 13 Gyr, and
  they are supplemented with the distribution at 13.5 Gyr. The
ordinate gives the age of the binary population in Gyr.}
\label{perdis}
\end{figure}

We made the following assumptions about the population of cool close
binaries in a GC.

\begin{enumerate}

\item The initial period distribution is constant for the period interval
  2.0-2.9 d and decreases linearly to zero at 1.4 d:
\begin{eqnarray}
n(P,M_1,q){\rm d}P & = &  0 \enskip {\rm{for}}\enskip P <
1.4\,,\nonumber \\
n(P,M_1,q){\rm d}P & = &  \frac{5}{3}(P-1.4)\enskip {\rm{for}}\enskip
1.4 \le P < 2\,,\\
n(P,M_1,q){\rm d}P & = & 1 \enskip {\rm{for}}\enskip P \ge 2\,,\nonumber
\label{initdistr}
\end{eqnarray}

where $n(P,M_1,q)$ is the number density of binaries with different
periods and mass ratios.{ The value of 1.4 results from a linear
  approximation of the step-wise distribution considered here, where a
  first non-zero value of $n$ is for $P = 1.5$ d. The lack of shorter
  initial periods is due to the inflated component radii during the T
  Tauri phase (see Sect. 2.2).} 

\item The initial period and component masses are not correlated,
  that is, \thinspace  
      $n(P,M_1,q)dPdMdq$ = $n_1(P)n_2(M_1)n_3(q)dPdMdq$, where 
      $n_1(P), n_2(M_1)$ and 
   $n_3(q)$ are the number densities depending on period, mass, and mass ratio,
   respectively.{ This follows from the semi-empirically determined
   properties   of
   the birth binary population \citep{krou95}.}

\item The primary mass distribution is given by the initial mass function of
   \citet{sal55}:  $n_2(M_1) \propto M_1^{-2.35}$. { Alternative forms of
   this function have also been proposed by
   \citet{mill79} and \citet{sca86}. More recently, \citet{krou02}
   confirmed the Salpeter mass function for stars with masses higher
   than 0.5 $M_{\sun}$ , except that he suggested a somewhat flatter mass
   dependence with an exponent of 2.3. Nevertheless, the original
   Salpeter relation is still in use. The difference
   between both exponents has a negligible influence on the mass
   function over the narrow mass interval considered here.}

\item The mass ratio distribution is flat, $n_3(q)$ =
  const.{ \citep{rag10,mil12}, although short-period field binaries may
  show some preference to $q \approx 1$ \citep{duch13}. In lack of
  detailed information on the mass ratio distribution of GC binaries, we 
adopted the flat distribution.}
\end{enumerate}

\begin{table*}
\caption{Model binaries with periods shorter than 1 d at the age of 11 Gyr}
\label{bin11}
\centering 
\begin{tabular}{lllllrll}
\hline
\hline
model  & $M_1(M_{\sun})$  &  $M_2(M_{\sun})$ & $q$ & $P$(d)  & $\log
(L/L_{\sun})$  &  $\log T_e$ & Remarks\\
\hline
0.75+0.4(1.6) &   0.458 & 0.618 & 0.741 & 0.168 & -0.489 & 3.723 & CB,MS\\
0.75+0.5(1.5) &   0.526 & 0.643 & 0.818 & 0.177 & -0.380 & 3.731 & CB,MS\\
0.75+0.7(1.5) &   0.580 & 0.764 & 0.759 & 0.194 & -0.149 & 3.757 & CB,MS\\
0.8+0.4(1.8) &    0.441 & 0.678 & 0.650 & 0.173 & -0.363 & 3.737 & CB,MS\\
0.8+0.5(1.7) &    0.544 & 0.667 & 0.816 & 0.226 & -0.295 & 3.737 & CB,MS\\
0.8+0.6(1.6) &    0.547 & 0.752 & 0.727 & 0.220 & -0.154 & 3.754 & CB,MS\\
0.8+0.7(1.6) &    0.535 & 0.850 & 0.629 & 0.213 &  0.017 & 3.777 & CB,MS\\
0.81+0.5(1.7) &   0.487 & 0.732 & 0.665 & 0.201 & -0.223 & 3.750 & CB,MS\\
0.81+0.6(1.6) &   0.489 & 0.818 & 0.598 & 0.198 & -0.057 & 3.770 & CB,MS\\
0.82+0.4(1.9) &   0.447 & 0.690 & 0.648 & 0.183 & -0.327 & 3.740 & CB,MS\\
0.82+0.5(1.7) &   0.460 & 0.767 & 0.600 & 0.166 & -0.166 & 3.759 & CB,MS\\
0.82+0.7(1.7) &   0.659 & 0.742 & 0.888 & 0.283 & -0.085 & 3.751 & CB,MS\\
0.83+0.4(1.9) &   0.432 & 0.713 & 0.606 & 0.145 & -0.284 & 3.745 & CB,MS\\
0.83+0.5(1.8) &   0.525 & 0.710 & 0.739 & 0.222 & -0.217 & 3.746 & CB,MS\\
0.83+0.6(1.7) &   0.552 & 0.771 & 0.716 & 0.240 & -0.101 & 3.758 & CB,MS\\
0.83+0.7(1.7) &   0.584 & 0.827 & 0.706 & 0.256 &  0.008 & 3.769 & CB,MS\\
0.84+0.4(2.0) &   0.442 & 0.711 & 0.622 & 0.193 & -0.282 & 3.745 & CB,MS\\
0.84+0.5(1.8) &   0.451 & 0.792 & 0.569 & 0.205 & -0.122 & 3.765 & CB,MS\\
0.84+0.6(1.7) &   0.481 & 0.852 & 0.565 & 0.219 &  0.011 & 3.778 & CB,MS\\
0.84+0.7(1.7) &   0.521 & 0.898 & 0.580 & 0.229 &  0.117 & 3.788 & CB,MS\\
0.85+0.5(1.9) &   0.542 & 0.710 & 0.763 & 0.237 & -0.192 & 3.746 & CB,MS\\
0.85+0.6(1.8) &   0.611 & 0.731 & 0.836 & 0.271 & -0.108 & 3.750 & CB,MS\\
0.86+0.4(2.1) &   0.415 & 0.755 & 0.550 & 0.190 & -0.217 & 3.755 & CB,MS\\
0.86+0.5(1.9) &   0.452 & 0.808 & 0.559 & 0.214 & -0.086 & 3.769 & CB,MS\\
0.86+0.6(1.8) &   0.507 & 0.843 & 0.601 & 0.244 &  0.010 & 3.775 & CB,MS\\
0.86+0.7(1.8) &   0.589 & 0.847 & 0.695 & 0.275 &  0.055 & 3.772 & CB,MS\\
0.865+0.4(2.2) &  0.474 & 0.700 & 0.677 & 0.226 & -0.261 & 3.743 & CB,MS\\
0.865+0.5(1.9) &  0.444 & 0.821 & 0.541 & 0.176 & -0.055 & 3.772 & CB,MS\\
0.865+0.5(2.0) &  0.556 & 0.709 & 0.784 & 0.261 & -0.172 & 3.746 & CB,MS\\
0.865+0.6(1.8) &  0.478 & 0.875 & 0.546 & 0.214 &  0.063 & 3.784 & CB,MS\\
0.865+0.7(1.8) &  0.527 & 0.913 & 0.577 & 0.260 &  0.148 & 3.790 & CB,MS\\
0.865+0.8(1.9) &  0.662 & 0.863 & 0.767 & 0.305 &  0.117 & 3.773 & CB,MS\\
0.8675+0.4(2.2) & 0.378 & 0.797 & 0.474 & 0.242 & -0.097 & 3.767 & CB,MS\\
0.8675+0.4(2.3) & 0.432 & 0.744 & 0.581 & 0.292 & -0.123 & 3.752 & CB,MS\\
0.8675+0.5(2.0) & 0.420 & 0.846 & 0.496 & 0.276 &  0.025 & 3.776 & CB,MS\\
0.8675+0.5(2.1) & 0.508 & 0.758 & 0.670 & 0.345 & -0.023 & 3.754 & CB,MS\\
0.8675+0.6(1.9) & 0.475 & 0.879 & 0.540 & 0.321 &  0.123 & 3.780 & CB,MS\\
0.8675+0.6(2.0) & 0.595 & 0.759 & 0.784 & 0.412 &  0.079 & 3.754 & CB,MS\\
0.8675+0.7(1.8) & 0.426 & 1.012 & 0.421 & 0.242 &  0.336 & 3.818 & CB,BS\\
0.8675+0.7(1.9) & 0.537 & 0.903 & 0.595 & 0.372 &  0.204 & 3.781 & CB,MS\\
0.8675+0.8(1.9) & 0.479 & 1.046 & 0.458 & 0.331 &  0.403 & 3.821 & CB,BS\\
0.8675+0.8(2.0) & 0.593 & 0.934 & 0.635 & 0.489 &  0.318 & 3.780 & CB,MS\\
0.88+0.5(2.0) &   0.191 & 1.072 & 0.178 & 0.845 &  0.543 & 3.823 & NCB,BS\\
0.9+0.4(2.4) &    0.189 & 1.001 & 0.189 & 0.843 &  0.436 & 3.802 & NCB,MS\\
0.9+0.6(1.9) &    0.152 & 1.221 & 0.124 & 0.709 &  0.836 & 3.887 & NCB,BS\\
\hline
\end{tabular}
\end{table*}    

{ Before calculating the period distribution,
  a weight for each CCBM must be determined to be used in the binary summation.} 
The initial CCBMs are distributed uniformly in $P,$ which means that
each model represents binaries from an equal period interval. When
summation is performed over consecutive period bins, the weight $w_P$,
attached to each model, is therefore constant for $P \ge 2$ d and decreases
linearly to zero, as given by Eq.~12.  The situation is
more complicated when we consider weights $w_M$ and $w_q$ connected with the initial
primary mass and initial mass ratio, because the models are distributed
nonuniformly in $M_1$ and $q$. To obtain their values, it was
necessary first to calculate the widths of
the respective intervals in $M_1$ and $q$, centered on each model, and then
to multiply them by factors resulting from assumptions 3 and 4 above. 
Finally, a product of all three weights was attached to each
of the 975 CCBMs and kept constant over all evolutionary calculations. 

Figure~\ref{perdis} presents the time evolution of the period
distribution with the 0.1 d bin size. The evolving distribution is shown
every 1 Gyr, starting from the initial distribution at the bottom of
the figure untill the age of 13 Gyr. The last
distribution, at the top of the figure, corresponds to the age of
13.5 Gyr. Broken lines show zero lines of the consecutive
distributions.  

The main evolutionary effects of the period distribution consist of
the shift of the short-period limit toward the shorter periods, together
with a stretching or flattening of the { distribution, except for its long-period
  end, where a deficit of binaries is visible. The deficit is due to
  the period increase of several binaries beyond the three-day limit, which is not
  compensated for by the possible period shortening of binaries with initial
  periods $\ge$ 3 d. They were not included in the present models,
  but we can estimate their approximate influence on the plotted
  distribution. All binaries with initial periods 3-3.5 d shorten
  them below 3 d during phase I, but those with primaries more
  massive than the TO mass quickly increase their periods above 3 d past RLOF
  (similarly as shown in Fig.~\ref{P0.88}). Lower mass binaries with
  the same initial periods remain detached over the Hubble time and
  shorten their periods continuously, but lose only about 15-20\% of the
  initial AM because the AML rate is inversely proportional to the period,
  see Eq.~\ref{amloss}. Their final periods fall between 2.4-3 d,  so they would partly fill in the deficit if they are included in
  the distribution.{ Nontheless, a comparison between the
    observed and predicted distribution should be limited to periods
    shorter than 2.4 d.} Binaries with initial periods above 3.5 never
  enter the period interval plotted in Fig.~\ref{perdis}. This
means that   the distribution does not go to zero at the long-period boundary, but
  its exact shape there depends on the initial period
  distribution beyond 3 d, which is not considered.

  The general properties of the period distribution are
  insensitive to the bin size, for example, increasing the bin size
  twice results  in a less ragged shape, but the broadening
  toward short periods and flattening of the distribution are still
  clearly visible.}

The short-period limit crosses the 1 d mark at the age of 4 Gyr and
0.5 d mark at about 7 Gyr. This is a very interesting result. It shows
that the value of the short-period limit of detached MS binaries 
in a stellar cluster can be used as an
independent measure of the cluster age, similarly as the rotation
period of cool MS stars is a measure of stellar age
\citep{bar09,mei11}. At present, very few clusters are known with a
sufficiently complete sample of binary stars to safely determine a
short-period limit. However, the increasing accuracy of photometric
observations in the future will enable detecting light
variations caused by the ellipticity of noneclipsing binary components. As a
result, a high percentage of all close binaries will be detected in
several clusters and a short-period limit of the period distribution
can be accurately determined. { Since close binaries can also be formed by
collisions with other cluster members (ignored
here), their
  possible effects on this limit will have to be quantified before the
  method is applied to a particular cluster.}

The flattening of the distribution with time is accompanied by a
decrease of the total number of binaries within the considered period
interval that is due to two mechanisms: merging of the two components, and a period
increase beyond 3 d of Algol-type binaries. As a result, the total
weighted number of CCBMs (equivalent to the integral of the
distribution over period) in the period interval 0.15-3.0 d decreases
by{ about one third (more accurately, by 32.9\%)} at 13.5 Gyr,
compared to the initial data. Unweighted number of models included
in the distribution decreased at that time from an initial 975 binaries to 398 { (41\%).}

We conclude that the predicted{ shape of the} period distribution
of cool close binaries at the age of a GC is flat{ over the
  interval  0.1-2.5 d} and bears no similarity to the initial
distribution.

\subsection{Mergers and contact binaries in a globular cluster}

As we showed above, the binary evolution past RLOF depends on the
relative importance of slow mass transfer that lengthens the orbital
period and of the AML that shortens it. Binaries with rapidly increasing period{ remain in} an Algol configuration, whereas the rest experience a
contact phase with the ultimate merging of the components{ if the
  timescale for a merger is shorter than the cluster age}.

\begin{table}
\caption{BS mergers formed between 10 and 11 Gyr, evolved to the age of 11 Gyr}
\label{merg11}
\centering 
\begin{tabular}{lllll}
\hline
\hline
model  & $t_{\mathrm{merg}}$(years)  &  $M(M_{\sun})$ & $\log
(L/L_{\sun})$  &  $\log T_e$\\
\hline
0.75+0.4(1.5) &   1.037E+10 & 0.981 & 0.331 & 3.841\\
0.8+0.4(1.7) &    1.040E+10 & 1.024 & 0.467 & 3.857\\
0.8+0.5(1.6) &    1.075E+10 & 1.114 & 0.657 & 3.888\\
0.8+0.6(1.5) &    1.042E+10 & 1.204 & 0.971 & 3.912\\
0.8+0.7(1.5) &    1.011E+10 & 1.294 & 1.072 & 3.906\\
0.81+0.4(1.7) &   1.005E+10 & 1.033 & 0.529 & 3.862\\
0.81+0.4(1.8)   & 1.092E+10 & 1.028 & 0.439 & 3.856\\
0.81+0.5(1.6) &   1.044E+10 & 1.122 & 0.728 & 3.891\\
0.81+0.6(1.5) &   1.014E+10 & 1.216 & 1.141 & 3.793\\
0.82+0.4(1.8) &   1.057E+10 & 1.039 & 0.505 & 3.862\\
0.82+0.5(1.6) &   1.014E+10 & 1.134 & 0.846 & 3.894\\
0.82+0.6(1.6) &   1.086E+10 & 1.217 & 0.856 & 3.910\\
0.83+0.4(1.8) &   1.018E+10 & 1.051 & 0.572 & 3.869\\
0.83+0.5(1.7) &   1.072E+10 & 1.138 & 0.734 & 3.895\\
0.83+0.6(1.6)   & 1.055E+10 & 1.229 & 0.995 & 3.913\\
0.84+0.4(1.9) &   1.059E+10 & 1.056 & 0.547 & 3.869\\
0.84+0.5(1.7) &   1.038E+10 & 1.149 & 0.830 & 3.898\\
0.84+0.6(1.6) &   1.024E+10 & 1.241 & 1.251 & 3.748\\
0.85+0.4(1.9)   & 1.019E+10 & 1.068 & 0.625 & 3.875\\
0.85+0.4(2.0) &   1.099E+10 & 1.061 & 0.528 & 3.868\\
0.85+0.5(1.7) &   1.002E+10 & 1.160 & 1.003 & 3.863\\
0.85+0.5(1.8) &   1.087E+10 & 1.153 & 0.749 & 3.899\\
0.85+0.6(1.7) &   1.085E+10 & 1.242 & 0.911 & 3.912\\
0.85+0.7(1.7) &   1.085E+10 & 1.330 & 1.048 & 3.921\\
0.86+0.4(2.0) &   1.049E+10 & 1.074 & 0.615 & 3.876\\
0.86+0.5(1.8) &   1.046E+10 & 1.165 & 0.867 & 3.905\\
0.86+0.6(1.7) &   1.045E+10 & 1.254 & 1.128 & 3.831\\
0.86+0.7(1.7) &   1.053E+10 & 1.340 & 1.341 & 3.733\\
0.865+0.4(2.0) &  1.023E+10 & 1.080 & 0.672 & 3.879\\
0.865+0.4(2.1) &  1.099E+10 & 1.074 & 0.566 & 3.874\\
0.865+0.5(1.8) &  1.021E+10 & 1.172 & 0.987 & 3.885\\
0.865+0.6(1.7) &  1.022E+10 & 1.262 & 1.133 & 3.829\\
0.865+0.8(1.8) &  1.064E+10 & 1.429 & 1.802 & 3.716\\
0.8675+0.4(2.0) & 1.015E+10 & 1.083 & 0.728 & 3.881\\
0.8675+0.4(2.1) & 1.090E+10 & 1.077 & 0.637 & 3.878\\
0.8675+0.5(1.8) & 1.011E+10 & 1.173 & 1.017 & 3.773\\
0.8675+0.5(1.9) & 1.086E+10 & 1.168 & 0.827 & 3.897\\
0.8675+0.6(1.8) & 1.093E+10 & 1.254 & 0.983 & 3.916\\
0.87+0.4(2.0) &   1.005E+10 & 1.086 & 0.772 & 3.880\\
0.87+0.5(1.9) &   1.080E+10 & 1.168 & 0.897 & 3.908\\
0.88+0.4(2.1) &   1.031E+10 & 1.092 & 0.822 & 3.878\\
0.88+0.5(1.9)   & 1.033E+10 & 1.181 & 1.029 & 3.760\\
0.88+0.6(1.8) &   1.041E+10 & 1.271 & 1.275 & 3.739\\
0.9+0.4(2.2) &    1.011E+10 & 1.110 & 0.986 & 3.808\\
\hline
\end{tabular}
\end{table}

The model calculations show that coalescence typically occurs in less
than 1 Gyr after RLOF, and the typical lifetime of a merger in the
region of BS is also of the order of 1 Gyr. We therefore focused
on binaries with a RLOF occurring at, or later than, 10 Gyr,
which then evolve toward a CB or NCB configuration. Binaries evolving
to the Algol configuration are not discussed.

\begin{figure}
\includegraphics[height=\hsize]{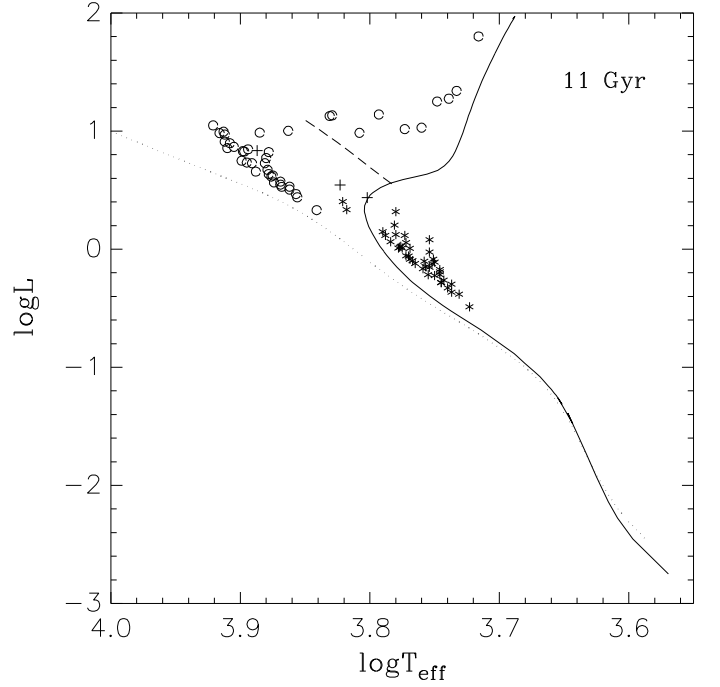}
\caption{HR diagram with model CBs, NCBs, and mergers at the age of 11
  Gyr, together with the corresponding low-metal isochrone. Asterisks
  denote binaries with an orbital period shorter than 0.5 d, plus signs
  binaries with periods between 0.5 and 1 d, and open circles denote
  mergers formed past 10 Gyr and evolved to the age of 11 Gyr.{
    The ZAMS is marked with a dotted line and the adopted red boundary
  of the BS region with long dashes.}}
\label{gc11}
\end{figure}

We selected model binaries in which RLOF occurred between 10 and 11
Gyr and which have orbital periods shorter than 1 d at the
age of 11 Gyr. In many of them (particularly those with the shortest
periods), both components already fill their Roche lobes, but in others
only the loser does, whereas the gainer is within its lobe.
The computed stellar
radii result strictly from the evolutionary models without allowing
for any effects that might influence them (e.\thinspace
g., magnetic fields or deviations from thermal equilibrium). The
observed radii of cool, rapidly rotating stars are often larger than
the model radii, so several real stars with the same global parameters
as CCBMs may in fact be already CBs even if they look like NCBs in our simulations. We marked binaries with $P < 0.5$ d as CBs
and those with $1 > P \ge 0.5$ d as NCBs and indicate them with asterisks
and plus signs, respectively. Numerical data on these models are given
in Table~\ref{bin11} with the consecutive columns giving the name of
the model, component masses, mass ratio, period, luminosity, effective
temperature, and remarks: CB - contact binary, NCB - near-contact
binary, MS - binary lies below TO, BS - binary lies in the BS
region. { The same data are plotted in Fig.~\ref{gc11} together
  with the ZAMS and a 11 Gyr isochrone. In addition, the red boundary of the
  BS region we adopted is shown as the
  long-dashed line. This is the ZAMS line shifted upward by
  $\Delta\log L = 0.8$ (= 2 mag). It lies close to the
  TAMS line for the most massive BSs and somewhat higher for TO stars
  \citep{sch92}. A similar red boundary was adopted by \citet{lei07}
  and \citet{lei11a}, but \citet{fio14}, for instance, used a vertical
  line in the HRD to delimit the BS region.}


Forty-five binaries { (4.6\% of all models)} fulfill these
conditions. Three of them have periods between 0.5 and 1 d, and the
rest has periods shorter than 0.5 d.  All CBs and NCBs are clustered
near the TO point. However, most of them lie below TO, with a
mere four falling
in the region of BS. The situation is different for binaries with $P
> 1$ d (not plotted). They all are of Algol-type with low mass ratios
and lie in the BS region{ or redward of it}.  It is interesting
to note that progenitors of CBs and NCBs have initial periods from a
narrow interval: 33 out of 45 had periods between 1.7-2.0 d and none
had a period longer than 2.4 d. Their total initial masses are between
1.15-1.5 $M_{\sun}$ , but they are lowered to 1.08-1.37 $M_{\sun}$ at
the age of 11 Gyr. Short periods dominate among the plotted binaries:
apart from three stars with periods longer than 0.5 d (plus signs), there
are only seven more with periods between 0.3-0.5 d and the rest,
that is, 35 stars, has periods shorter than 0.3 d, with the
record shortest equal to $\sim 0.14$ d. Similarly short periods
dominate among field CBs, although here the shortest known period is
close to 0.2 d \citep{ruc07}. We can expect periods shorter
than 0.2 d among low-metallicity CBs, however, because their components are more
compact than their solar-metallicity counterparts.  Binaries with
periods shorter than 0.3 d have a very high mass ratio, between
0.5-0.8, whereas{ those with longer periods have lower ratios down
  to 0.1-0.2. This trend results from the dependence of the mass ratio
  past mass exchange on the difference in core helium content between
  the components. For two chemically uniform ZAMS stars with unequal
  masses, the mass transfer stops at $q = 1$ \citet{kui41}. The
  difference in the core helium content between a more and less
  massive component increases in the course of evolution, so that the later
  mass transfer takes place, the lower the mass ratio. Observations of field CBs with periods shorter than 0.3 d
  indicate that most of them belong to least massive CBs with both
  components still in the early phases of the MS life
  \citep{steg12}. Their mass ratios cluster around 0.5, as opposed to
  longer period CBs \citep{ruc10}.} Most CBs with periods shorter
than 0.3 d stay in contact for less than 1 Gyr, and then coalescence
occurs as a consequence of the outer Roche lobe overflow. The longer
period binaries live longer as CBs or NCBs \citep[up to 2 Gyr, see
also][]{tian06} before merging as a result of the Darwin instability
\citep{rasio95}.

\begin{figure}
\includegraphics[height=\hsize]{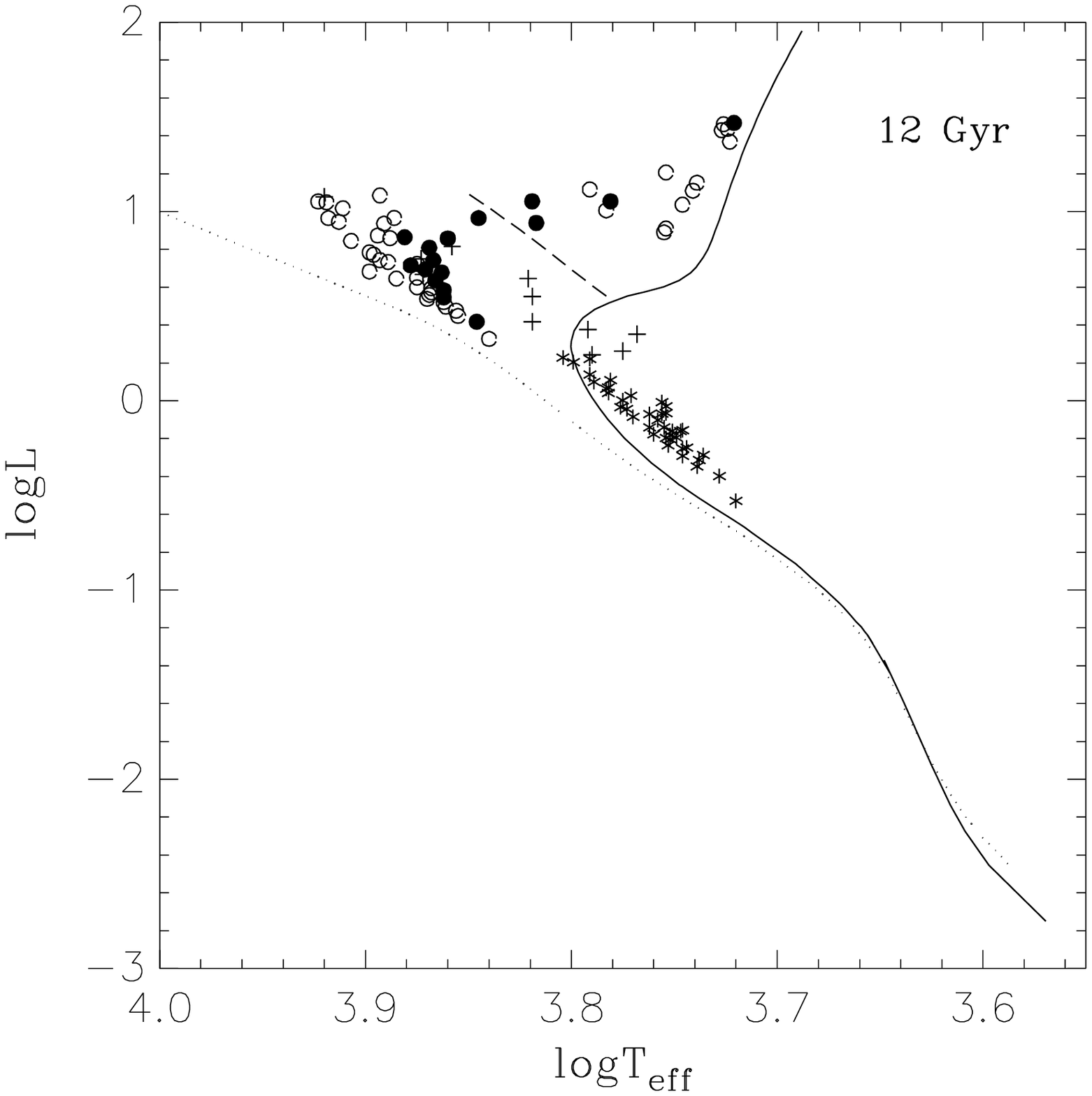}
\caption{HR diagram with model CBs, NCBs, and mergers at the age of 12
  Gyr, together with the corresponding low-metal isochrone. Asterisks
  denote binaries with an orbital period shorter than 0.5 d, plus signs
  binaries with periods between 0.5 and 1 d, filled circles denote
  mergers formed between 10-11 Gyr, and open circles those formed
  between 11-12 Gyr. All mergers are evolved to the age of 12 Gyr. The{
    ZAMS is marked with a dotted line and the adopted red boundary
  of the BS region with long dashes.}}
\label{gc12}
\end{figure}

\begin{figure}
\includegraphics[height=\hsize]{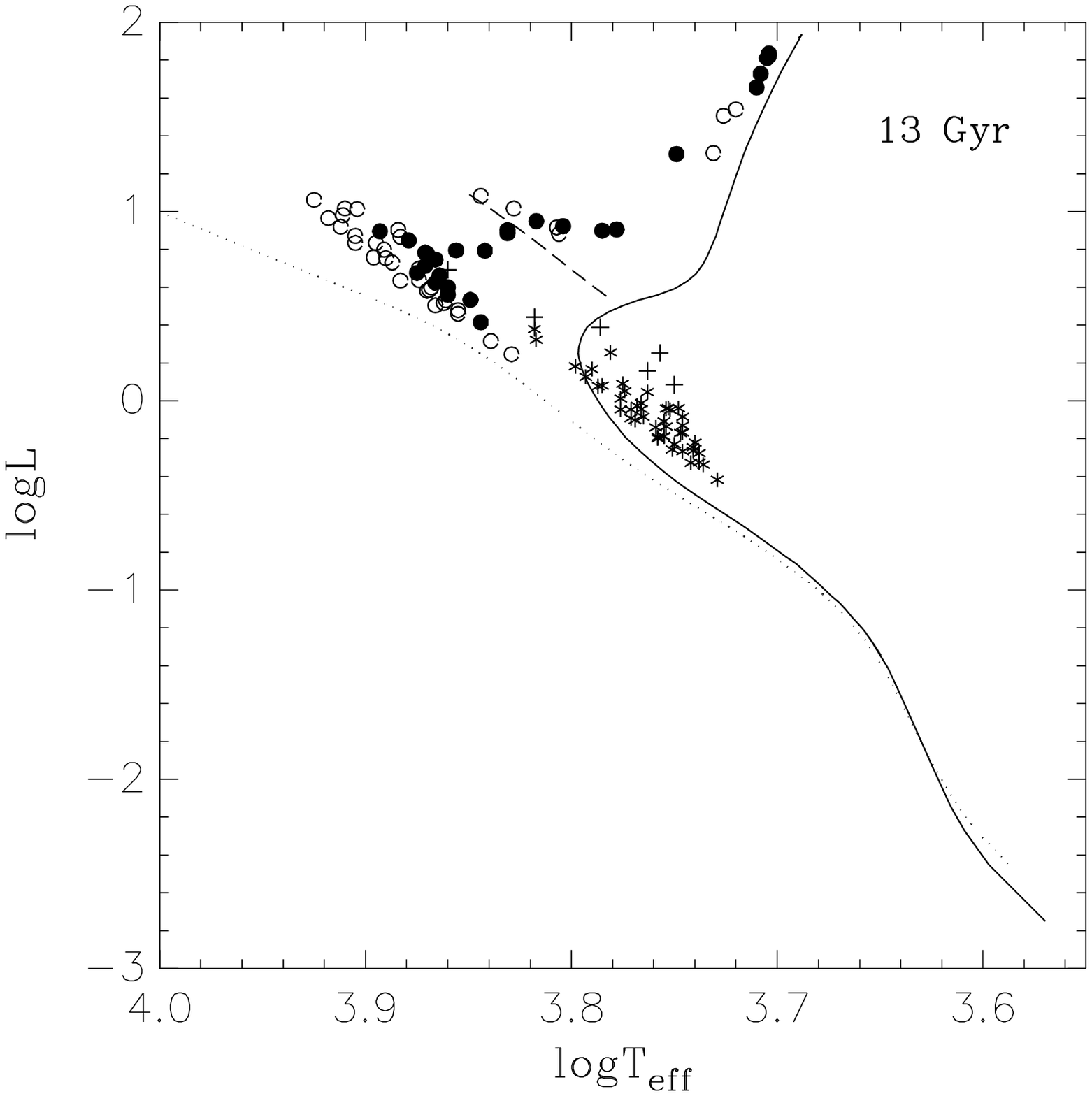}
\caption{HR diagram with model CBs, NCBs, and mergers at the age of 13
  Gyr, together with the corresponding low-metal isochrone. Asterisks
  denote binaries with an orbital period shorter than 0.5 d, plus signs
  binaries with periods between 0.5 and 1 d, filled circles denote
  mergers formed between 10-12 Gyr, and open circles those formed
  between 12-13 Gyr. All mergers are evolved to the age of 13 Gyr.{
    The ZAMS is marked with a dotted line and the adopted red boundary
  of the BS region with long dashes.}}
\label{gc13}
\end{figure}

\begin{table*}
\caption{Model binaries with periods shorter than 1 d at the age of 12 Gyr}
\label{bin12}
\centering 
\begin{tabular}{lllllrll}
\hline
\hline
model  & $M_1(M_{\sun})$  &  $M_2(M_{\sun})$ & $q$ & $P$(d)  & $\log
(L/L_{\sun})$  &  $\log T_e$ & Remarks\\
\hline
0.7+0.4(1.5) &    0.445 & 0.583 & 0.763 & 0.142 & -0.528 & 3.720 & CB,MS\\
0.75+0.5(1.6) &   0.541 & 0.619 & 0.874 & 0.199 & -0.398 & 3.728 & CB,MS\\
0.75+0.7(1.6) &   0.603 & 0.731 & 0.825 & 0.223 & -0.194 & 3.749 & CB,MS\\
0.8+0.4(1.9) &    0.426 & 0.686 & 0.621 & 0.167 & -0.348 & 3.739 & CB,MS\\
0.8+0.5(1.8) &    0.545 & 0.657 & 0.830 & 0.229 & -0.287 & 3.736 & CB,MS\\
0.8+0.6(1.7) &    0.552 & 0.738 & 0.748 & 0.229 & -0.161 & 3.751 & CB,MS\\
0.8+0.7(1.7)   &  0.533 & 0.842 & 0.633 & 0.231 &  0.002 & 3.775 & CB,MS\\
0.81+0.5(1.8) &   0.468 & 0.743 & 0.630 & 0.206 & -0.203 & 3.752 & CB,MS\\
0.81+0.6(1.7) &   0.471 & 0.827 & 0.570 & 0.215 & -0.044 & 3.773 & CB,MS\\
0.81+0.7(1.7) &   0.511 & 0.872 & 0.586 & 0.187 &  0.070 & 3.782 & CB,MS\\
0.82+0.4(2.0) &   0.410 & 0.718 & 0.571 & 0.183 & -0.290 & 3.746 & CB,MS\\
0.82+0.5(1.8) &   0.439 & 0.780 & 0.563 & 0.162 & -0.141 & 3.762 & CB,MS\\
0.82+0.6(1.8) &   0.602 & 0.705 & 0.854 & 0.280 & -0.155 & 3.746 & CB,MS\\
0.82+0.7(1.8) &   0.634 & 0.759 & 0.835 & 0.295 & -0.066 & 3.754 & CB,MS\\
0.83+0.4(2.1) &   0.458 & 0.679 & 0.675 & 0.208 & -0.312 & 3.738 & CB,MS\\
0.83+0.5(1.9) &   0.486 & 0.742 & 0.655 & 0.223 & -0.183 & 3.752 & CB,MS\\
0.83+0.6(1.8) &   0.526 & 0.788 & 0.668 & 0.248 & -0.071 & 3.762 & CB,MS\\
0.83+0.7(1.8) &   0.565 & 0.836 & 0.676 & 0.264 &  0.026 & 3.771 & CB,MS\\
0.84+0.4(2.1) &   0.398 & 0.747 & 0.533 & 0.183 & -0.236 & 3.753 & CB,MS\\
0.84+0.5(1.9) &   0.424 & 0.811 & 0.523 & 0.198 & -0.086 & 3.770 & CB,MS\\
0.84+0.6(1.8) &   0.457 & 0.866 & 0.528 & 0.220 &  0.041 & 3.782 & CB,MS\\
0.84+0.7(1.8) &   0.498 & 0.911 & 0.547 & 0.236 &  0.138 & 3.791 & CB,MS\\
0.85+0.4(2.2) &   0.435 & 0.717 & 0.607 & 0.209 & -0.252 & 3.746 & CB,MS\\
0.85+0.5(2.0) &   0.487 & 0.756 & 0.644 & 0.239 & -0.141 & 3.755 & CB,MS\\
0.85+0.6(1.9) &   0.565 & 0.766 & 0.738 & 0.273 & -0.067 & 3.756 & CB,MS\\
0.85+0.7(1.9) &   0.650 & 0.769 & 0.845 & 0.312 & -0.007 & 3.756 & CB,MS\\
0.86+0.4(2.2) &   0.391 & 0.772 & 0.506 & 0.166 & -0.178 & 3.760 & CB,MS\\
0.86+0.4(2.3) &   0.459 & 0.704 & 0.652 & 0.228 & -0.248 & 3.744 & CB,MS\\
0.86+0.5(2.0) &   0.420 & 0.834 & 0.504 & 0.200 & -0.035 & 3.776 & CB,MS\\
0.86+0.5(2.1) &   0.537 & 0.714 & 0.752 & 0.277 & -0.159 & 3.747 & CB,MS\\
0.86+0.6(1.9) &   0.466 & 0.873 & 0.534 & 0.248 &  0.064 & 3.783 & CB,MS\\
0.86+0.7(1.9) &   0.543 & 0.883 & 0.615 & 0.279 &  0.109 & 3.781 & CB,MS\\
0.865+0.4(2.3) &  0.413 & 0.753 & 0.548 & 0.214 & -0.197 & 3.754 & CB,MS\\
0.865+0.5(2.1) &  0.485 & 0.771 & 0.629 & 0.256 & -0.103 & 3.758 & CB,MS\\
0.865+0.6(1.9) &  0.452 & 0.891 & 0.507 & 0.196 &  0.100 & 3.789 & CB,MS\\
0.865+0.6(2.0) &  0.587 & 0.755 & 0.777 & 0.314 & -0.030 & 3.754 & CB,MS\\
0.865+0.7(1.9) &  0.485 & 0.944 & 0.514 & 0.267 &  0.204 & 3.799 & CB,MS\\
0.865+0.8(2.0) &  0.574 & 0.939 & 0.611 & 0.311 &  0.221 & 3.791 & CB,MS\\
0.8675+0.4(2.3) & 0.215 & 0.952 & 0.226 & 0.388 &  0.229 & 3.804 & CB,MS\\
0.8675+0.4(2.4) & 0.233 & 0.931 & 0.250 & 0.553 &  0.244 & 3.790 & NCB,MS\\
0.8675+0.4(2.5) & 0.275 & 0.887 & 0.310 & 0.705 &  0.262 & 3.775 & NCB,MS\\
0.8675+0.4(2.6) & 0.283 & 0.877 & 0.323 & 0.955 &  0.352 & 3.768 & NCB,MS\\
0.8675+0.5(2.1) & 0.221 & 1.032 & 0.214 & 0.583 &  0.418 & 3.819 & NCB,BS\\
0.8675+0.5(2.2) & 0.279 & 0.972 & 0.287 & 0.711 &  0.377 & 3.792 & NCB,MS\\
0.8675+0.6(1.9) & 0.179 & 1.161 & 0.154 & 0.579 &  0.668 & 3.872 & NCB,BS\\
0.8675+0.6(2.0) & 0.256 & 1.080 & 0.237 & 0.790 &  0.552 & 3.819 & NCB,BS\\
0.8675+0.7(1.9) & 0.224 & 1.199 & 0.187 & 0.655 &  0.750 & 3.873 & NCB,BS\\
0.8675+0.7(2.0) & 0.296 & 1.126 & 0.263 & 0.924 &  0.645 & 3.821 & NCB,BS\\
0.8675+0.8(1.9) & 0.148 & 1.363 & 0.109 & 0.630 &  1.078 & 3.920 & NCB,BS\\
0.8675+0.8(2.0) & 0.275 & 1.229 & 0.224 & 0.940 &  0.816 & 3.858 & NCB,BS\\
\hline
\end{tabular}
\end{table*}

In addition to binaries,{ 44 mergers (4.5\% of all models)},
formed between 10 and 11 Gyr and evolved to the age of 11 Gyr, are
listed in Table~\ref{merg11} and plotted in Fig.~\ref{gc11} as open
circles. The columns give the name of the model, age of merging, merger
mass, luminosity, and effective temperature. In addition to 34 stars lying in
the BS region, 10 subgiants or giants are also shown for comparison. {
  These are mergers formed shortly after 10 Gyr, which already
left the MS
  and now approach the red giant branch.} Because the orbital AM of a
binary just before merging is far too high to be accommodated in a
spin AM of a single star, most of it must be expelled together with
some amount of mass in the form of a ring or disk surrounding the
freshly formed star{ \citep{ras95}}. We adopted 0.1 $M_{\sun}$ for
the expelled mass, based on an indirect argument presented by
\citet{stekir13b}.{ This mass ejected to the distance of
  10-20~$R_{\sun}$ can absorb all excess AM
  \citep{sil01,stekir13b}. The resulting merger mass is equal to the
  sum of the component masses minus 0.1~$M_{\sun}$}.  Mergers are
assumed to evolve like single stars with the same degree of hydrogen
depletion \citep[i.\thinspace e., no mixing of the merger matter is
assumed, see][]{sil09}.


Disregarding subgiants, mergers form a well-defined
sequence that lies close to the ZAMS for the lowest masses and
increasingly deviates from it for the increasing mass. The most
massive merger with a mass of 1.43 $M_{\sun}$ also has the highest
luminosity and already climbs the red giant branch
(Fig.~\ref{gc11}). All other mergers have masses between 0.98-1.33
$M_{\sun}$.

{ How does the situation change with age? Figure~\ref{gc12} shows
  similar data as Fig.~\ref{gc11}, but 1 Gyr later. Here, again, all
  binaries with periods shorter than 1 d at the age of 12 Gyr are
  plotted with the same coding as in Fig.~\ref{gc11}. Most binaries
  with periods shorter than 0.5 d are new because almost all those
  plotted in Fig.~\ref{gc11} did not survive till 12 Gyr and merged in
  the meantime. But seven binaries from Fig.~\ref{gc11} survived for more
  than 1 Gyr and are also plotted in Fig.~\ref{gc12}. Mergers formed
  between 11-12 Gyr are shown as open circles. They are supplemented
  with the mergers from the previous figure, which did not yet climb
  the red giant branch. Those mergers are shown as filled circles. The
  numerical data are given in Tables~\ref{bin12} and \ref{merg12}.}

The CBs and NCBs present at the age of 12 Gyr have similar properties as
those existing 1 Gyr earlier. Almost all short-period binaries lie
below or very close to the TO point. Out of 39 binaries { (4\% of all
  modeled)}, only one lies slightly above TO, but out of 11 long-period
binaries 7 lie in the region of BS (two are hidden in a dense clump
of filled circles). Similarly as in the previous case, most of the
initial orbital periods come from a narrow interval of 1.8-2.1 d that is
shifted longward by 0.1 d compared to the binaries plotted in
Fig.~\ref{gc11}. The total initial masses lie between 1.1-1.67 $M_{\sun}$
, but they are lowered to 1.03-1.5 $M_{\sun}$ at 12 Gyr.

{ Of the mergers formed between 11-12 Gyr (open circles in
  Fig.~\ref{gc12}), 30 (3\% of the total number of CCBMs) lie in the
  BS region and 11 (1.1\% ) are subgiants. An additional 16
  mergers { (1.6\%)} were formed before 11 Gyr (filled
  circles). Twelve of these still lie in the BS region, which
  indicates that a significant fraction of BS spends more than 1 Gyr
  in this region. Three of them still remain in the BS region even
  at the age of 13 Gyr, that is, almost 3 Gyr after formation
  (see below).}

\begin{table}
\caption{BS mergers formed between 10 and 12 Gyr, evolved to the age of 12 Gyr}
\label{merg12}
\centering 
\begin{tabular}{lllll}
\hline
\hline
model  & $t_{\mathrm{merg}}$(years)  &  $M(M_{\sun})$ & $\log
(L/L_{\sun})$  &  $\log T_e$\\
\hline
0.75+0.4(1.5) &  1.037E+10 & 0.981 & 0.418 & 3.846\\
0.8+0.4(1.7) &   1.040E+10 & 1.024 & 0.584 & 3.862\\
0.8+0.5(1.6) &   1.075E+10 & 1.114 & 0.864 & 3.881\\
0.81+0.4(1.7) &  1.005E+10 & 1.033 & 0.678 & 3.863\\
0.81+0.4(1.8) &  1.092E+10 & 1.028 & 0.546 & 3.862\\
0.82+0.4(1.8) &  1.057E+10 & 1.039 & 0.635 & 3.866\\
0.83+0.4(1.8) &  1.018E+10 & 1.051 & 0.743 & 3.867\\
0.84+0.4(1.9) &  1.059E+10 & 1.056 & 0.694 & 3.871\\
0.85+0.5(1.8) &  1.087E+10 & 1.153 & 1.054 & 3.781\\
0.86+0.4(2.0) &  1.049E+10 & 1.074 & 0.809 & 3.869\\
0.865+0.4(2.0) & 1.023E+10 & 1.080 & 0.940 & 3.817\\
0.865+0.4(2.1) & 1.099E+10 & 1.074 & 0.715 & 3.878\\
0.8675+0.4(2.0) &1.015E+10 & 1.083 & 1.468 & 3.721\\
0.8675+0.4(2.1) &1.090E+10 & 1.077 & 0.858 & 3.860\\
0.87+0.4(2.0) &  1.005E+10 & 1.086 & 0.965 & 3.845\\
0.87+0.5(1.9) &  1.080E+10 & 1.168 & 1.054 & 3.819\\
0.75+0.4(1.6) &  1.146E+10 & 0.974 & 0.327 & 3.840\\
0.75+0.5(1.5) &  1.143E+10 & 1.065 & 0.538 & 3.870\\
0.75+0.6(1.5) &  1.194E+10 & 1.148 & 0.683 & 3.898\\
0.75+0.7(1.5) &  1.126E+10 & 1.242 & 1.085 & 3.893\\
0.8+0.4(1.8) &   1.133E+10 & 1.016 & 0.476 & 3.856\\
0.8+0.5(1.7) &   1.178E+10 & 1.105 & 0.646 & 3.885\\
0.8+0.6(1.6) &   1.149E+10 & 1.195 & 0.945 & 3.913\\
0.81+0.4(1.9) &  1.183E+10 & 1.021 & 0.448 & 3.855\\
0.81+0.5(1.7) &  1.143E+10 & 1.116 & 0.733 & 3.889\\
0.81+0.6(1.6) &  1.117E+10 & 1.206 & 1.117 & 3.791\\
0.82+0.4(1.9) &  1.141E+10 & 1.034 & 0.521 & 3.862\\
0.82+0.5(1.7) &  1.108E+10 & 1.127 & 0.859 & 3.888\\
0.82+0.6(1.7) &  1.188E+10 & 1.208 & 0.845 & 3.907\\
0.82+0.7(1.7) &  1.173E+10 & 1.294 & 1.053 & 3.923\\
0.83+0.4(1.9) &  1.103E+10 & 1.045 & 0.595 & 3.868\\
0.83+0.4(2.0) &  1.187E+10 & 1.038 & 0.498 & 3.861\\
0.83+0.5(1.8) &  1.167E+10 & 1.130 & 0.743 & 3.893\\
0.83+0.6(1.7) &  1.154E+10 & 1.219 & 1.017 & 3.911\\
0.83+0.7(1.7) &  1.144E+10 & 1.307 & 1.461 & 3.726\\
0.84+0.4(2.0) &  1.144E+10 & 1.050 & 0.573 & 3.868\\
0.84+0.5(1.8) &  1.129E+10 & 1.141 & 0.873 & 3.894\\
0.84+0.6(1.7) &  1.120E+10 & 1.230 & 1.436 & 3.724\\
0.85+0.4(2.1) &  1.180E+10 & 1.056 & 0.559 & 3.869\\
0.85+0.5(1.9) &  1.177E+10 & 1.146 & 0.772 & 3.896\\
0.85+0.6(1.8) &  1.178E+10 & 1.234 & 0.965 & 3.918\\
0.85+0.7(1.8) &  1.183E+10 & 1.319 & 1.049 & 3.919\\
0.86+0.4(2.1) &  1.127E+10 & 1.068 & 0.651 & 3.875\\
0.86+0.5(1.9) &  1.130E+10 & 1.158 & 0.936 & 3.891\\
0.86+0.6(1.8) &  1.137E+10 & 1.246 & 1.206 & 3.754\\
0.86+0.7(1.8) &  1.149E+10 & 1.331 & 1.430 & 3.727\\
0.865+0.4(2.2) & 1.178E+10 & 1.069 & 0.600 & 3.875\\
0.865+0.5(1.9) & 1.105E+10 & 1.164 & 1.006 & 3.783\\
0.865+0.5(2.0) & 1.189E+10 & 1.157 & 0.784 & 3.898\\
0.8675+0.4(2.2)& 1.158E+10 & 1.070 & 0.723 & 3.875\\
0.8675+0.5(2.0)& 1.158E+10 & 1.161 & 0.966 & 3.886\\
0.87+0.4(2.1) &  1.148E+10 & 1.068 & 1.369 & 3.723\\
0.87+0.4(2.2) &  1.179E+10 & 1.055 & 0.910 & 3.754\\
0.87+0.6(1.8) &  1.162E+10 & 1.238 & 0.890 & 3.755\\
0.88+0.4(2.2) &  1.151E+10 & 1.069 & 1.152 & 3.739\\
0.89+0.4(2.2) &  1.123E+10 & 1.088 & 1.036 & 3.746\\
0.89+0.4(2.3) &  1.144E+10 & 1.076 & 1.110 & 3.741\\
\hline
\end{tabular}
\end{table}

The situation after another 1 Gyr is displayed in Fig.~\ref{gc13} with
the same coding as in Figs.~\ref{gc11}-\ref{gc12}. The numerical data of
the plotted models are given in Tables~\ref{bin13} and \ref{merg13}.
We have here 46 short-period { (4.7\%) and 6 long-period  (0.6\%)}
CBs and NCBs. Only four binaries (two short-period and two long-period
binaries)
lie above the 13 Gyr isochrone in the region of BS, and all others
lie below the TO point. Similarly as in the earlier age, the initial
periods of most of the binaries come from a narrow range 1.9-2.2
d, shifted again longward by 0.1 d, compared to the age of 12 Gyr. The
initial total masses of the binaries are between 1.2-1.67 $M_{\sun}$
, but the present masses are lowered to 1.0-1.5 $M_{\sun}$.

Thirty-six mergers { (3.7\%)}, formed past 12 Gyr, are plotted in
Fig.~\ref{gc13} as open circles. They are supplemented with 24 mergers
{ (2.5\%)} formed between 11 and 12 Gyr and 3 more, formed shortly
after 10 Gyr. Fifteen mergers fall in the subgiant region, and the
rest lies in the  BS region, among them the three{ low-mass}binaries
 mentioned
above. Merger masses do not differ from the masses of the mergers plotted
in the previous figures: they fall into the interval of 0.93-1.32
$M_{\sun}$.

\section{Comparison of model calculations with 
observations}

\subsection{Contact and near-contact binaries}

Recently, results of a deep search for photometric variables in a few
GCs have been published \citep{kath09,kaetal10,kaetal13a,karo13}. In
all cases, no CBs were detected among unevolved cluster members,
although a prominent binary sequence is visible above the cluster MS
on the CMD \citep[see e.\thinspace g.,][]{kaetal13a}. The authors
concluded that CBs begin only to appear about 1 mag below the
TO. Our
results are fully consistent with this conclusion. Assuming that TO
lies approximately at $\log L = 0$ in Figs.~\ref{gc11}-\ref{gc13}, the
faintest CBs lie about 0.4-0.5 in $\log L$ below TO, which corresponds
to 1-1.2 mag. All model binaries with still lower luminosity are
detached. The complete lack of faint, low-mass model CBs results from the
adopted value of 1.5 d for the initial orbital cut-off
period. Lowering this value (e.\thinspace g., to 1 d or less) leads to
a contact configuration reached within the age of a GC by binaries
with essentially unevolved, low-mass components.  Their absence among the
observed CBs is direct evidence that the cut-off
period is not shorter than 1.5 d. It may even be close to 2 d
\citep{ste11a}. Adopting the latter value would result in a
disappearance of several of the faintest CBs in
Figs.~\ref{gc11}-\ref{gc13} (and some of the plotted mergers).

\begin{table*}
\caption{Model binaries with periods shorter than 1 d at the age of 13 Gyr}
\label{bin13}
\centering 
\begin{tabular}{lllllrll}
\hline
\hline
model  & $M_1(M_{\sun})$  &  $M_2(M_{\sun})$ & $q$ & $P$(d)  & $\log
(L/L_{\sun})$  &  $\log T_e$ & Remarks\\
\hline
0.7+0.5(1.5) &    0.480 & 0.633 & 0.758 & 0.142 & -0.418 & 3.729 & CB,MS\\
0.7+0.6(1.5) &    0.518 & 0.680 & 0.762 & 0.178 & -0.326 & 3.738 & CB,MS\\
0.75+0.6(1.6) &   0.509 & 0.732 & 0.695 & 0.185 & -0.230 & 3.750 & CB,MS\\
0.8+0.4(2.0) &    0.408 & 0.699 & 0.584 & 0.167 & -0.328 & 3.742 & CB,MS\\
0.8+0.5(1.9) &    0.528 & 0.667 & 0.792 & 0.236 & -0.276 & 3.738 & CB,MS\\
0.8+0.6(1.8) &    0.565 & 0.716 & 0.789 & 0.241 & -0.171 & 3.747 & CB,MS\\
0.8+0.7(1.8) &    0.551 & 0.816 & 0.675 & 0.245 & -0.026 & 3.768 & CB,MS\\
0.81+0.4(2.1) &   0.446 & 0.668 & 0.668 & 0.199 & -0.338 & 3.736 & CB,MS\\
0.81+0.5(1.9) &   0.450 & 0.754 & 0.597 & 0.210 & -0.186 & 3.755 & CB,MS\\
0.81+0.6(1.8) &   0.468 & 0.822 & 0.569 & 0.225 & -0.047 & 3.771 & CB,MS\\
0.81+0.7(1.8) &   0.493 & 0.881 & 0.560 & 0.210 &  0.080 & 3.785 & CB,MS\\
0.82+0.4(2.1) &   0.386 & 0.737 & 0.524 & 0.179 & -0.256 & 3.751 & CB,MS\\
0.82+0.5(1.9) &   0.410 & 0.802 & 0.511 & 0.170 & -0.101 & 3.769 & CB,MS\\
0.82+0.5(2.0) &   0.537 & 0.675 & 0.796 & 0.253 & -0.220 & 3.740 & CB,MS\\
0.82+0.6(1.9) &   0.589 & 0.709 & 0.831 & 0.287 & -0.132 & 3.746 & CB,MS\\
0.82+0.7(1.9) &   0.636 & 0.747 & 0.851 & 0.307 & -0.048 & 3.752 & CB,MS\\
0.83+0.4(2.2) &   0.413 & 0.717 & 0.576 & 0.204 & -0.266 & 3.746 & CB,MS\\
0.83+0.5(2.0) &   0.448 & 0.771 & 0.581 & 0.227 & -0.141 & 3.759 & CB,MS\\
0.83+0.6(1.9) &   0.500 & 0.806 & 0.620 & 0.255 & -0.043 & 3.766 & CB,MS\\
0.83+0.7(1.9) &   0.541 & 0.849 & 0.637 & 0.277 &  0.051 & 3.774 & CB,MS\\
0.84+0.4(2.2) &   0.375 & 0.763 & 0.491 & 0.163 & -0.199 & 3.758 & CB,MS\\
0.84+0.4(2.3) &   0.449 & 0.690 & 0.651 & 0.228 & -0.264 & 3.741 & CB,MS\\
0.84+0.5(2.0) &   0.398 & 0.830 & 0.480 & 0.192 & -0.045 & 3.776 & CB,MS\\
0.84+0.5(2.1) &   0.521 & 0.707 & 0.737 & 0.263 & -0.168 & 3.746 & CB,MS\\
0.84+0.6(1.9) &   0.429 & 0.884 & 0.485 & 0.222 &  0.077 & 3.787 & CB,MS\\
0.84+0.6(2.0) &   0.611 & 0.704 & 0.868 & 0.324 & -0.084 & 3.746 & CB,MS\\
0.84+0.7(1.9) &   0.465 & 0.934 & 0.498 & 0.249 &  0.182 & 3.798 & CB,MS\\
0.85+0.4(2.3) &   0.383 & 0.764 & 0.501 & 0.204 & -0.191 & 3.758 & CB,MS\\
0.85+0.4(2.4) &   0.461 & 0.685 & 0.673 & 0.265 & -0.246 & 3.741 & CB,MS\\
0.85+0.5(2.1) &   0.438 & 0.795 & 0.551 & 0.240 & -0.085 & 3.765 & CB,MS\\
0.85+0.6(2.0) &   0.514 & 0.809 & 0.635 & 0.281 & -0.011 & 3.766 & CB,MS\\
0.85+0.7(2.0) &   0.600 & 0.808 & 0.743 & 0.322 &  0.047 & 3.763 & CB,MS\\
0.86+0.4(2.4) &   0.402 & 0.753 & 0.534 & 0.221 & -0.190 & 3.755 & CB,MS\\
0.86+0.5(2.2) &   0.486 & 0.756 & 0.643 & 0.270 & -0.109 & 3.755 & CB,MS\\
0.86+0.6(2.0) &   0.422 & 0.908 & 0.465 & 0.250 &  0.127 & 3.793 & CB,MS\\
0.86+0.6(2.1) &   0.612 & 0.717 & 0.854 & 0.333 & -0.038 & 3.748 & CB,MS\\
0.86+0.7(2.0) &   0.498 & 0.916 & 0.544 & 0.289 &  0.168 & 3.790 & CB,MS\\
0.865+0.4(2.4) &  0.346 & 0.811 & 0.427 & 0.220 & -0.092 & 3.771 & CB,MS\\
0.865+0.4(2.5) &  0.405 & 0.753 & 0.538 & 0.274 & -0.137 & 3.754 & CB,MS\\
0.865+0.4(2.6) &  0.398 & 0.760 & 0.524 & 0.414 & -0.041 & 3.754 & CB,MS\\
0.865+0.5(2.2) &  0.403 & 0.844 & 0.477 & 0.267 &  0.012 & 3.776 & CB,MS\\
0.865+0.5(2.3) &  0.496 & 0.750 & 0.661 & 0.342 & -0.039 & 3.753 & CB,MS\\
0.865+0.5(2.4) &  0.523 & 0.725 & 0.721 & 0.524 &  0.084 & 3.750 & NCB,MS\\
0.865+0.6(2.1) &  0.478 & 0.857 & 0.558 & 0.328 &  0.091 & 3.775 & CB,MS\\
0.865+0.6(2.2) &  0.508 & 0.827 & 0.614 & 0.516 &  0.158 & 3.763 & NCB,MS\\
0.865+0.7(2.0) &  0.411 & 1.007 & 0.408 & 0.281 &  0.322 & 3.817 & CB,BS\\
0.865+0.7(2.1) &  0.501 & 0.916 & 0.547 & 0.451 &  0.254 & 3.781 & CB,MS\\
0.865+0.8(2.1) &  0.470 & 1.031 & 0.456 & 0.330 &  0.378 & 3.818 & CB,BS\\
0.865+0.7(2.2) &  0.627 & 0.792 & 0.792 & 0.617 &  0.253 & 3.757 & NCB,MS\\
0.865+0.8(2.2) &  0.527 & 0.975 & 0.541 & 0.584 &  0.389 & 3.786 & NCB,MS\\
0.8675+0.4(2.4) & 0.125 & 1.023 & 0.122 & 0.806 &  0.443 & 3.818 & NCB,BS\\
0.8675+0.5(2.1) & 0.104 & 1.133 & 0.092 & 0.875 &  0.691 & 3.860 & NCB,BS\\
\hline
\end{tabular}
\end{table*}

On the whole, of 975 CCBMs, only about 50 { ($\sim$ 5\%)} are
in a contact phase at a given age (more accurately, there are 45, 50,
and 52 at the age of 11, 12, and 13 Gyr, respectively). Approximately
10\% of these{ CBs} fall in the BS region. Qualitatively similar
proportions were obtained by \citet{tian06}. They modeled over one
million binaries, and only less than 2000 were in the mass-exchange
phase at the age of 4 Gyr, corresponding to the age of M67. About 11\%
of these fell in the BS region. A more quantitative comparison between
their and our results cannot be made because of distinctly different
assumptions adopted.

A lower{ CB frequency in GCs than in the field} has already
been noticed by \citet{ruc00}. It results most likely from a very low
binary frequency of { $\sim$ 0.1 in GCs (but with a significant
  scatter among different clusters), compared to 0.5} in the field
\citep{mil12}, together with a low percentage of binaries in the
contact phase at a given age, as shown above. Any statistical
comparison of models with observations is therefore extremely uncertain because
we have so few observational data. Nevertheless, we can compare the predicted
to observed fraction of BSs among all CBs in a GC. We used data for two
clusters observed by Kaluzny and his collaborators with the highest
number of CBs (except for $\omega$ Centauri). In M4, \citet{kaetal13a}
detected nine CBs with one BS among them, and in 47 Tuc, \citet{karo13}
identified 15 CBs or NCBs, of which six are BS. The resulting fractions
are 0.1 and 0.4, respectively. We did not take into account $\omega$
Cen with the richest population of CBs \citep{kalo04} because it is
highly atypical and the accurate number of member CBs is not well known,
although the approximate data indicate a similarly high ratio as in case
of 47 Tuc. The predicted fractions range from 0.08 to 0.2 for the
three considered cluster ages.  As we see, they are close to the lower
observed value but are at odds with the higher value. This may
indicate some deficiencies of the CCBM, for instance, too short initial cut-off
periods, a too high AML rate at short periods, or a too low mass-transfer
rate. With the lower AML rate and/or higher mass transfer rate, a
binary stays longer in contact and can reach a lower mass ratio before
merging. Many of the short-period CBs have a rather high mass ratio of
about 0.7-0.8 at the time of merging, whereas field CBs with $P \la
0.3$ d center around a value of 0.5 \citep{ruc10,steg12}. The lower
mass ratio means more mass transferred to the { gainer}, hence its
higher mass and higher position on the HRD, that is, a higher
probability for entering the BS region. There may also still be another
explanation of the discrepancy: strong fluctuations of this fraction
among different clusters suggest a nonuniform BSs formation rate in
some clusters, with individual bursts occurring in the recent past
(see below). Regardless of the reason, it is apparent that profound
differences occur among GCs, which makes a comparison of
theoretical predictions with individual clusters uncertain.

\begin{table*}
\caption{BS mergers formed between 10 and 13 Gyr, evolved to the age of 13 Gyr}
\label{merg13}
\centering 
\begin{tabular}{llllllllll}
\hline
\hline
model  & $t_{\mathrm{merg}}$(years)  &  $M(M_{\sun})$ & $\log
(L/L_{\sun})$  &  $\log T_e$ &\qquad model  & $t_{\mathrm{merg}}$(years) 
&  $M(M_{\sun})$ & $\log(L/L_{\sun})$  &  $\log T_e$\\
\hline
0.75+0.4(1.5) &   1.037E+10 & 0.981 & 0.533 & 3.849 &\qquad 0.8+0.4(1.9) &    1.225E+10 & 1.010 & 0.479 & 3.855\\
0.8+0.4(1.7) &    1.040E+10 & 1.024 & 0.793 & 3.842 &\qquad 0.8+0.5(1.8) &    1.286E+10 & 1.096 & 0.635 & 3.883\\

0.85+0.5(1.8) &   1.087E+10 & 1.153 & 0.899 & 3.785 &\qquad 0.8+0.6(1.7) &    1.260E+10 & 1.185 & 0.872 & 3.905\\

0.75+0.4(1.6) &   1.146E+10 & 0.974 & 0.415 & 3.844 &\qquad 0.8+0.7(1.7) &    1.229E+10 & 1.272 & 1.506 & 3.726\\

0.75+0.5(1.5) &   1.143E+10 & 1.065 & 0.676 & 3.875 &\qquad 0.81+0.4(2.0) &   1.274E+10 & 1.016 & 0.460 & 3.855\\

0.75+0.6(1.5) &   1.194E+10 & 1.148 & 0.896 & 3.893 &\qquad 0.81+0.5(1.8) &   1.244E+10 & 1.107 & 0.731 & 3.887\\

0.8+0.4(1.8) &    1.133E+10 & 1.016 & 0.600 & 3.860 &\qquad 0.81+0.6(1.7) &   1.224E+10 & 1.196 & 1.082 & 3.844\\

0.8+0.5(1.7) &    1.178E+10 & 1.105 & 0.847 & 3.879 &\qquad 0.82+0.4(2.0) &   1.231E+10 & 1.026 & 0.532 & 3.861\\

0.81+0.4(1.9) &   1.183E+10 & 1.021 & 0.560 & 3.860 &\qquad 0.82+0.5(1.8) &   1.203E+10 & 1.119 & 0.867 & 3.883\\

0.81+0.5(1.7) &   1.143E+10 & 1.116 & 0.906 & 3.778 &\qquad 0.82+0.6(1.8) &   1.292E+10 & 1.200 & 0.834 & 3.905\\

0.82+0.4(1.9) &   1.141E+10 & 1.034 & 0.662 & 3.864 &\qquad 0.82+0.7(1.8) &   1.278E+10 & 1.285 & 1.016 & 3.910\\

0.83+0.4(1.9) &   1.103E+10 & 1.045 & 0.795 & 3.856 &\qquad 0.83+0.4(2.1) &   1.271E+10 & 1.032 & 0.517 & 3.862\\

0.83+0.4(2.0) &   1.187E+10 & 1.038 & 0.624 & 3.866 &\qquad 0.83+0.5(1.9) &   1.260E+10 & 1.123 & 0.754 & 3.890\\

0.83+0.5(1.8) &   1.167E+10 & 1.130 & 1.304 & 3.749 &\qquad 0.83+0.6(1.8) &   1.255E+10 & 1.209 & 1.013 & 3.904\\

0.84+0.4(2.0) &   1.144E+10 & 1.050 & 0.746 & 3.866 &\qquad 0.83+0.7(1.8) &   1.247E+10 & 1.297 & 1.309 & 3.731\\

0.85+0.4(2.1) &   1.180E+10 & 1.056 & 0.714 & 3.871 &\qquad 0.84+0.4(2.1) &   1.223E+10 & 1.044 & 0.599 & 3.868\\

0.85+0.5(1.9) &   1.177E+10 & 1.146 & 0.923 & 3.804 &\qquad 0.84+0.5(1.9) &   1.219E+10 & 1.134 & 0.903 & 3.884\\

0.86+0.4(2.1) &   1.127E+10 & 1.068 & 0.902 & 3.831 &\qquad 0.85+0.4(2.2) &   1.260E+10 & 1.049 & 0.585 & 3.869\\

0.86+0.5(1.9) &   1.130E+10 & 1.158 & 0.885 & 3.831 &\qquad 0.85+0.5(2.0) &   1.265E+10 & 1.137 & 0.798 & 3.891\\

0.865+0.4(2.2) &  1.178E+10 & 1.069 & 0.783 & 3.871 &\qquad 0.85+0.6(1.9) &   1.274E+10 & 1.225 & 0.981 & 3.911\\

0.865+0.5(1.9) &  1.105E+10 & 1.164 & 0.775 & 3.870 &\qquad 0.85+0.7(1.9) &   1.281E+10 & 1.309 & 1.062 & 3.925\\

0.865+0.5(2.0) &  1.189E+10 & 1.157 & 0.949 & 3.817 &\qquad 0.86+0.4(2.2) &   1.208E+10 & 1.062 & 0.698 & 3.874\\

0.8675+0.4(2.2) & 1.158E+10 & 1.070 & 1.835 & 3.704 &\qquad 0.86+0.4(2.3) &   1.282E+10 & 1.056 & 0.581 & 3.870\\

0.8675+0.5(2.0) & 1.158E+10 & 1.161 & 1.728 & 3.708 &\qquad 0.86+0.5(2.0) &   1.215E+10 & 1.152 & 1.016 & 3.828\\

0.87+0.4(2.1) &   1.148E+10 & 1.068 & 1.811 & 3.705 &\qquad 0.86+0.5(2.1) &   1.300E+10 & 1.143 & 0.758 & 3.896\\

0.87+0.4(2.2) &   1.179E+10 & 1.055 & 1.656 & 3.710 &\qquad 0.86+0.6(1.9) &   1.230E+10 & 1.236 & 1.540 & 3.720\\

0.88+0.4(2.2) &   1.151E+10 & 1.069 & 1.823 & 3.704 &\qquad 0.86+0.7(1.9) &   1.244E+10 & 1.321 & 2.126 & 3.700\\

0.89+0.4(2.2) &   1.123E+10 & 1.088 & 2.048 & 3.696 &\qquad 0.865+0.4(2.3) &  1.254E+10 & 1.062 & 0.638 & 3.874\\

0.7+0.4(1.5) &    1.226E+10 & 0.927 & 0.246 & 3.829 &\qquad 0.865+0.5(2.1) &  1.271E+10 & 1.151 & 0.834 & 3.895\\

0.75+0.4(1.7) &   1.261E+10 & 0.968 & 0.316 & 3.839 &\qquad 0.865+0.6(2.0) &  1.297E+10 & 1.233 & 0.965 & 3.918\\

0.75+0.5(1.6) &   1.269E+10 & 1.056 & 0.504 & 3.866 &\qquad 0.8675+0.4(2.3) & 1.276E+10 & 1.057 & 0.916 & 3.807\\

0.75+0.7(1.6) &   1.252E+10 & 1.230 & 0.919 & 3.912 &\qquad 0.8675+0.6(1.9) & 1.254E+10 & 1.233 & 0.881 & 3.806\\

\hline
\end{tabular}
\end{table*}

A direct comparison of CCBMs to the observed parameters of individual
CBs or NCBs is hampered because only very few{ high-quality} data are
available for these GC members.  To our knowledge, V60, an Algol-type
BS in M55, is the only NCB with parameters based on light and
velocity curves \citep{roz13}. Generally, to find a progenitor for a
given binary, a fine-tuning of the initial parameters is needed. Here
we do not attempt it, but we can still find a model that reproduces V60
quite well. This is the most massive CCBM, that is, 0.89+0.8(1.9). Its parameters at the age of about 11 Gyr are
compared with the observed values for V60 in
Table~\ref{v60}. The positions of the two binaries are also shown in the HRD
(Fig.~\ref{allbss}), where a diamond corresponds to V60 and the plus
sign to the model.  Another model, 0.9+0.8(1.9), evolves in about 11
Gyr to a similar configuration with a period of 1.18 d and
component masses of 0.24 and 1.30~$M_{\sun}$, which is also not far from
V60, but its luminosity, $\log L =$ 0.998, is somewhat too high.

\begin{table*}
\caption{Comparison of the observed parameters of
  binary V60 from M55 with one of CCBMs}
\label{v60}
\centering 
\begin{tabular}{lllllllll}
\hline
\hline
Parameter  & $P$ (d)  &  $a(R_{\sun})$ & $M_1 (M_{\sun})$ & $M_2
(M_{\sun})$ & $R_1 (R_{\sun})$ & $R_2 (R_{\sun})$ & $\log
(L/L_{\sun})$  &  $\log T_e$\\
\hline
V60 & 1.18 & 5.49$\pm$.05 & 0.33$\pm$.02 & 1.26$\pm$.03 & 1.48$\pm$.01 & 1.10$\pm$.02 & 0.813$\pm$.013 & 3.832$\pm$.008\\
model 0.89+0.8(1.9) & 1.18 & 5.41 & 0.28 & 1.26 & 1.42 & 1.13 & 0.880
& 3.854\\
\hline
\end{tabular}
\end{table*}

As we see, the model { parameters agree with the observed values
  within about $2\sigma$}, even though the model is not sufficiently
massive to better reproduce the observations.  The total mass of V60
is equal to 1.59 $M_{\sun}$, compared to
only 1.54 $M_{\sun}$ for the model. A slight mass increase,
together with a correct period and metallicity
should give a better agreement. When the initial
parameter values of CCBM were specified, the following evolution was
fully determined - we did not introduce any free parameters that
would, for instance, describe additional mass and/or AM loss.

{ The mass and radius of the less massive component of V60 indicate
  that it is an evolved star with a helium core of $\sim 0.15 M_{\sun}$ ,
  whereas its companion is a ZAMS star. The binary is thus a typical
  short-period Algol. It is commonly accepted that such a
  configuration results from a rapid mass transfer following a RLOF by
  an evolved primary. In effect, a binary is formed showing the well-known Algol paradox. No other way of producing an Algol-type
  binary seems to be viable, although many details of evolution
  preceding RLOF, starting from the origin of the initial binary,
  possible dynamical interactions with other stars, or amount of mass
  and AM lost during the consecutive evolutionary phases, are still
  unknown and need to be assumed for the calculated model. Even if all
  these details are fixed, as in case of the CCBM code, the results
  are equivocal. Different combinations of the initial parameters may
  result in the same final values, except that they have different age
  \citep{ste11b,piet12}. If the age of the modeled binary is not known,
  ambiguity remains. For a binary member of a GC the age is known,
which additionally constrains a possible initial model.}

Stellar parameters have also been obtained for a few more CBs and NCBs
in GCs, based solely on the photometric observations. These are NH
19, 30 and 31 in NGC 5466 \citep{kall92}, NJL 5 and V239 in $\omega$
Cen \citep{helt93,kai12}, V1 in NGC 6397 \citep{rub96}, V228 in 47 Tuc
\citep{kal07,sar08}, and V47 and V53 in M4 \citep{liu11}. In most
cases, however, the resulting binary parameters are very uncertain, so
any fine-tuning of the initial parameters to reproduce the obtained
results must await better observations. Still, if
the data are acceptable, we can always find a model close to the observed
variable. \citet{kai12} analyzed the photometry of V239,  an Algol-type
binary with $P = 1.19$ d, whose light curves were obtained
earlier by \citet{kalo04}. The solution indicates that the orbit
inclination is close to 90\degr, hence the resulting parameters are probably almost correct. The resulting component masses are $M_1 =
0.07$ and $M_2 = 1.20 M_{\sun}$. The model 0.8675+0.6(2.0) evolves
after 13.5 Gyr into a binary with a period of 1.36 d and component masses
0.10+1.20 $M_{\sun}$. We stopped the calculations at this age, but the mass
transfer was still ongoing and the period was decreasing, so after
the next few time steps the binary would resemble V239 even more. V228 in
47 Tuc is another  Algol-type star. Its present parameters are
$P$ = 1.15 d, $M_1$ = 0.20 and $M_2 = 1.51 M_{\sun}$ {
  \citep{kal07}}. \citet{sar08} modeled this star using his
evolutionary program. The program did not include magnetized winds
but, instead, it contained two free parameters individually adjusted
and describing the total mass and AM lost during the evolution. He
concluded that the progenitor binary had a period of 1.35 d and
component masses equal to 0.88 and 0.85 $M_{\sun}$.{ Allowing for
  mass loss by the winds and the present total mass of V228, we estimate that} the initial component masses must
have been close to 0.9 $M_{\sun}$. None of the CCBMs is that
massive. The metallicity of 47 Tuc is also several times higher than
used in CCBMs. However, the model 0.96+0.93(2.3) with the solar
composition, calculated for another purpose some time ago, evolves
after 12.4 Gyr into the configuration with period of 1.13 d and
component masses 0.20+1.49 $M_{\sun}$.  This is quite close to the
observed parameters of V228.

V47 and V53 are CBs with short periods of 0.27 and 0.31 d. The photometric solutions resulted in component masses
of 0.20+1.66 $M_{\sun}$ for V47 and 0.74+0.91 $M_{\sun}$ for V53 {
  \citep{liu11}}. Unfortunately, the orbit inclination resulting from
the photometric solution is equal to about 72\degr and 40\degr for V47
and V53, respectively, so quite far from 90\degr. In such cases the
mass ratio is poorly constrained and can be far from the spectroscopic
value \citep{ruc06}. The binary V47, with a total mass of 1.86
$M_{\sun}$, is substantially more massive than any of the CCBMs
discussed in this paper. In fact, the mass of the primary is higher
than twice the TO mass { (equal to 0.8 $M_{\sun}$,
  \citet{kaetal13a})}, which casts { some} doubt on its
determination,{ although another route of producing a massive
  merger is possible, involving multiple mergers or collisions
  \citep{berg01, sand03, leigh11}}. Clearly, a high mass of the primary is
also difficult to reconcile with its low luminosity $L_{\mathrm pr} =
2 L_{\sun}$ \citep{liu11}. The expected initial mass of another
binary, V53, is about 1.8 $M_{\sun}$, which is again beyond the limits
considered by us. Still, we can find models in the CCBM set with the
period and $q$ similar to V53, for instance, the model
0.8675+0.8(1.9), which evolves after 11 Gyr into a binary with a
period of 0.305 d and component masses of 0.66+0.86 $M_{\sun}$.  We
did not attempt to find { possible} progenitors of the binaries
investigated by \citet{kall92}, \citet{helt93}, and \citet{rub96}
because their parameters are even more uncertain.

A more accurate search for progenitors of individual binaries in GCs
will be the subject of a separate paper.

\subsection{Blue stragglers}

Recent progress in understanding the origin and properties of
BS in GCs has been enormous, mostly due to observations of GC cores
with the Hubble telescope of{ \thinspace 
  \citet{pio04}, \citet{lei07}, \citet{bec12}, \citet{dal13a}, \citet{dal13b},
  and \citet{san14}}.  A comprehensive review of 
observations and resulting conclusions can be found in \citet{bof14}.

To compare models with observations, we first need to specify which
members of a GC are classified as BS. There is no single generally
accepted method. To avoid contamination from photometric errors of MS
stars at the TO point and blends (particularly in a cluster core),
selection criteria require that a BS candidate be brighter than TO by
a predefined quantity. In addition, a BS should lie at the extension
of the cluster MS, but the width of the BS band is not strictly
determined and varies among different authors. Some authors plot boxes
in the color-magnitude diagram (CMD), see, for instance, \citet{lei11a}, \citet{dal13b}, \citet{li13}, and \citet{fio14}, others
select BS by analyzing individual candidates and applying specific
criteria \citep{mor08,fer09}. The resulting set of BS approximately
extends over 1-2 mag in $V$ and 0.4-0.5 in $B-V$.  { Most of the
  observed BSs lie along the extension of the MS, but some stars, lying already
  beyond the TAMS, where evolved single stars and binaries can be found,
  are often included in the BS samples. The BS region adopted in
this paper is bound by the ZAMS, the corresponding isochrone (11, 12, or
13 Gyr), and the ZAMS line shifted by $\Delta L$ = +0.8
(Figs.~\ref{gc11}-\ref{gc13}).}

\begin{figure}
\includegraphics[height=\hsize]{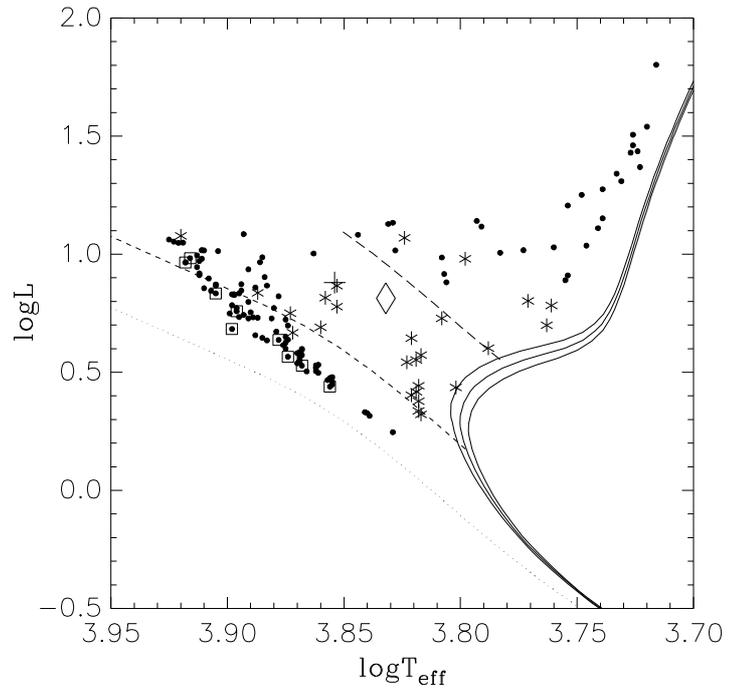}
\caption{Collective HR diagram of all mergers (filled circles) and
  all binary BS (asterisks) from Figs.~\ref{gc11}-\ref{gc13}, 
together with the corresponding isochrones. Rapidly
  rotating mergers are additionally marked with squares. The position
  of the eclipsing binary V60 from M55 and of the model 0.89+0.8(1.9)
  are plotted as a diamond and a plus sign. The ZAMS and the
  blue boundary of the binary BS region determined by \citet{tian06} are
  shown as a dotted and broken line.{ Long dashes
    mark the adopted red boundary of the BS region.}}
\label{allbss}
\end{figure}

\citet{fer09} detected a gap dividing BS in the cluster M30 into two
sequences: blue and red. They argued that the separation results from
two different formation mechanisms: the blue sequence, lying close to
the ZAMS, originated from collision-formed mergers, whereas the red
sequence that contains BS formed through mass transfer in binaries.
Similar sequences, albeit with a less pronounced gap between them,
were detected in NGC 362 \citep{dal13b}. { The blue sequence agrees
  with the expected position of mergers formed by a collision of two
  stars with masses close to half a TO mass. Such stars are hardly
  evolved before merging, hence a freshly formed merger is expected to
  lie close to ZAMS. The line separating both sequences was adopted
  after} \citet{tian06}. It lies approximately at the extension of the
MS binary sequence, that is, the ZAMS line shifted upward by
0.75 mag, and reproduces the expected position of unevolved,
equal-mass binaries.{ It follows from this interpretation} that the
blue sequence consists of single unevolved stars, whereas the red
sequence should consist of binaries. In addition, the distribution of
BS with a clear gap between both sequences requires that they were all
formed in a single { recent} burst, identified by \citet{fer09} with
a cluster core collapse. The subsequent evolution of blue BSs will
smear the distribution over CMD, so the gap should disappear. In addition, if
BSs are formed continuously in a cluster, none of the features observed by
\citet{fer09} and \citet{dal13b} should occur \citep{li13}. To see how
the models are situated relative to the blue and red sequences seen in
M30 and NGC 362, we plot all mergers from Tables~\ref{merg11},
\ref{merg12}, and \ref{merg13} as filled circles in Fig.~\ref{allbss}
and all binary BSs as asterisks. A few binaries with $1 < P < 3$ d are
also added. We do not list numerical data for them because the
accuracy of
the CCBM calculations decreases with increasing period of an
evolutionary advanced binary, therefore their parameters are rather
uncertain. The isochrones at 11, 12 and 13 Gyr are also plotted as solid
lines. In addition, the ZAMS line (dotted), the upper edge of the
binary BS region, determined by \citet{tian06} (broken line),{ and
  the adopted red boundary of the BS region (long dashes)} are
shown. The broken line delimits the region of binary BSs from
the blue side. All CCBMs but one lie to the right of this line.  The
exception is  model 0.8675+0.6(1.9). It became a CB at the age of
11 Gyr, but was lying below TO at that time (Table~\ref{bin11}). By
a continuing mass transfer, the gainer reached a mass of 1.161
$M_{\sun}$ at 12 Gyr and the binary moved to the blue part of the BS
region (Table~\ref{bin12}). The gainer is hardly evolved at that age
and the influence of the redder secondary is insufficient to move the
binary to the red part of the BS region. This indicates that in
exceptional cases CBs can also be found in the blue sequence, as is
observed in M30 and NGC 362 \citep{fer09,dal13b}.

{ In contrast to binaries, the model mergers can be found in both
  parts of the BS region (Fig.\ref{allbss}). Those with the lowest
  masses, around 1 $M_{\sun}$, lie close to ZAMS, along a narrow blue
  sequence. They were formed by two unevolved stars, hence resulting
  mergers are also unevolved even at the age of 13 Gyr.{ Their
    masses are between 0.9-1.1 $M_{\sun}$.} A gap
  separating them from the binaries is clearly visible. However,
  mergers more massive than about 1.1 $M_{\sun}$ are formed by two
  stars of which at least one has substantially depleted core hydrogen, which means that they are evolutionary advanced already immediately after
  formation. In effect, they { arrive to} the sequence tilted to ZAMS
  and extending to the red part of the BS region. Similarly, mergers
  formed collisionally in the evolutionary models of a GC are  scattered over the whole BS region \citep{sil09,sil13,chat13}. We
  conclude that binaries that move to the BS region as a result
of mass
  exchange, arrive predominantly to the red of the line
  determined by \citet{tian06}. However, the existence of a gap
  delimiting the narrow sequence of BSs close to ZAMS and extending
to the most massive BSs cannot be reconciled with a
  continuous merger production. A burst in their production seems to
  be the most plausible explanation, as suggested by
  \citet{fer09}, although this picture {\it per se} is not proof of
  the collisional formation of blue BSs.}

{
\begin{figure}
\includegraphics[height=\hsize]{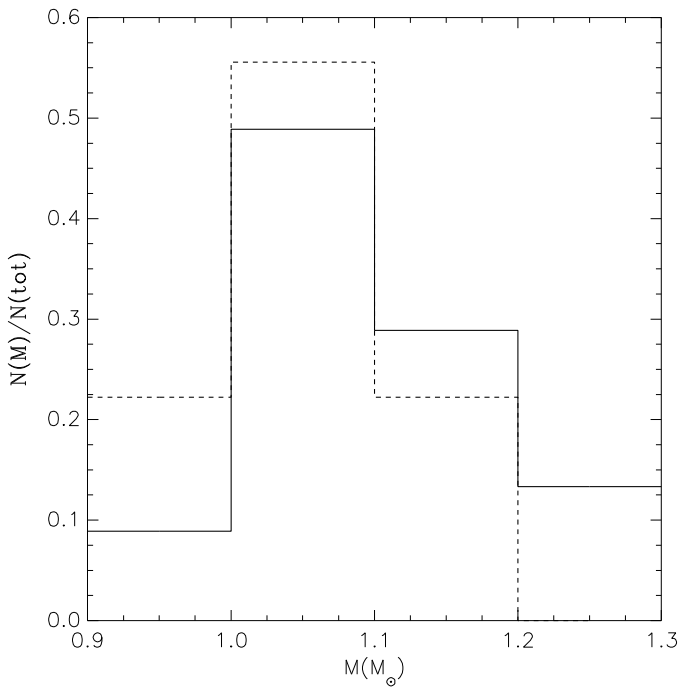}
\caption{Mass distribution of the BS mergers from Table~\ref{merg13}
  (solid line) compared to the observed distribution of the pulsation
  masses of SX Phe stars in NGC 6541 (broken line). Both
  distributions are normalized to unity.}
\label{massdistr}
\end{figure}}

Masses of a few binary BS are known from the solution of the
eclipsing light curves, sometimes supplemented with radial
velocity curves (see previous subsection). Less is known about masses
of single BS. Until recently, two methods have been applied to
determine
their masses. The first was based on spectral and photometric
observations, the second on the evolutionary tracks in the CMD {
  \citep{sha97,gel11,gel12,dem05,kal07,karo09,kaetal13a}}.
Unfortunately, discrepancies often occur when both methods are used on
the same object{ \citep{dem05,gel12}}. Recently, a new method was
introduced by \citet{fio14}. It is based on the comparison of
the pulsation properties of BS that belong to SX Phe - type
variables.{ Applying this method to SX Phe stars in NGC 6541
  resulted in the pulsation masses of nine variable BSs. Six of them
  are pulsating in the fundamental mode and three in the first
  overtone. For the last three stars the authors also calculated masses
  assuming fundamental pulsation, but they clearly preferred the first
  overtone identification. The masses agree very well with
  evolutionary masses, particularly when they are binned according to
  luminosity. We compare them with masses of the mergers from the
  BS region in Fig.~\ref{gc13}. There are 45 such mergers in
  Table~\ref{merg13}. Figure~\ref{massdistr} shows the mass
    distribution of the model mergers (solid line) together with the
    observed distribution (broken line). Both distributions are
    normalized to unity. The agreement is surprisingly good, taking
    into account the low number of the observed masses and the fact that
    SX Phe instability strip is narrower than the BS region. The lack
  of the observed masses in the last bin, between 1.2-1.3 $M_{\sun}$
   , most likely results from this restriction.} 

Recently, spectroscopic studies of BSs in a number of GCs have been
performed with the aim to investigate the mechanisms of their formation
\citep[][and references therein]{lov13,lov14}. One of the goals was to
measure rotation rates of BSs. Freshly formed binary mergers are
expected to rotate rapidly even though a great majority of the orbital
AM is expelled with { some matter \citep{ras95}}. Rapid rotation is
also expected for collision-formed mergers unless the collision is
strictly central{ \citep{sil97,sil01}}. But cool stars lose AM by
the magnetized wind. In effect, they spin down from an initial rotation
period of a fraction of a day to about one week in $10^8$ years,
that is, in terms of the equatorial velocity, from
$v_{\rm{eq}} \approx 100-150$ km/s to less than 10 km/s
\citep{bar09}. Except, possibly, for the most massive mergers, all
other BSs in GCs are magnetically active and spin down on a timescale
of ~$10^8$ years. In Fig.~\ref{allbss} we highlighted all mergers
formed in the last $10^8$ years before 11, 12, and 13 Gyr with squares.{ Rotation evolution is not included
  in the CCBM code, but we assume that freshly formed mergers rotate
  rapidly.} As we see, all these mergers lie close to the high-temperature boundary of the BS region, indicating that they are still
unevolved. There are only nine rapid rotators in the figure,
that is, about 10\% of all BS mergers.{ This is an upper limit for the
  expected fraction of rapid rotators in a GC, achieved only if
  mergers formed in the last Gyr exist in the BS region. All mergers
  formed earlier will add to slowly rotating, thus decreasing the
  fraction of rapid rotators. For example, only three such rotators are
  expected to occur among 48 BS mergers in Fig.~\ref{gc13} (see
  Table~\ref{merg13}), which is equivalent to about 6\%.}

So far, accurate observations of $v\sin i$ of several BSs in four GCs
have been obtained: 47 Tuc \citep{ferrar06}, M4 \citep{l10}, NGC 6397
\citep{l12}, and M30 \citep{lov13}. The total number of BSs with
a measured rotation rate in these clusters is 96, ten of which are rapid
rotators with $v\sin i > 50$ km/s (we exclude one W UMa-type binary
counted as a rapid rotator in M30), this is about 10\%. This
is close to the predicted value (see above). Moreover, all rapid
rotators in M4 and moderate rotators (with $v\sin i > 10$ km/s) in 47
Tuc lie near the blue boundary of the BS region, indicating that they
are freshly formed \citep{ferrar06,kaetal13a}. However,{
  this statistics is poorer when individual clusters are
  considered:} eight rapid rotators are present in M4 (40\% of the total
measured sample in this cluster), whereas only two are members of
three other clusters, of more than 80 stars measured. This shows again
that GCs are very different from each other, and a comparison of an
individual cluster with theoretical predictions { may be misleading.}

If mergers spin down as a result of magnetized winds, we expect a gap in
the rotation distribution between rapidly and slowly rotating stars,
resulting from the existence of the so-called Vaughan-Preston gap
\citep{bar11}. The rotation distributions of BSs show a gap
\citep[see Fig.~10 in][]{l12}, although its statistical significance is
low.

Another goal of the spectroscopic survey of BSs was to search for
anomalous surface abundances in BSs that would indicate the
material processed by the CNO cycle {
  \citep{ferrar06,l10,l12,lov13}}. As a star evolves, the
abundance of carbon $^{\rm{12}}$C and $^{\rm{16}}$O decreases, whereas
the abundance of nitrogen $^{\rm{14}}$N increases in the layers where
nuclear reactions operate \citep{iben67}.  The transition between the
cosmic and modified abundance moves with time slowly outward until it
reaches the mass fraction $\sim 0.3$ for solar-type stars
\citep{iben67}. If enough matter is transferred to the companion, the
material processed by nuclear reactions is exposed, hence anomalous
abundances of these elements { are expected to} be present on the surfaces of both
stars. Detection of any of these anomalies in a BS can be treated as
an evidence for the mass transfer channel of its
formation. An underabundance of carbon has indeed been observed in
field Algols \citep{sar96}.

It follows from these considerations that at least 70\% of the
loser mass needs to be transferred to its companion to expose altered
abundances{ \citep{iben67,sar96}}.  In most CCBMs, only about
40-50\% of the mass is transferred to the gainer before the two
components merge.{ We should therefore not expect to see any CNO
  peculiarities in these binaries and mergers. A sufficient mass
  fraction is transferred only in a few binaries that ultimately merge
  as a result of the Darwin instability}. There are about ten such BSs among
the binaries plotted in Fig.~\ref{allbss}. All except one lie in the
red part of the BS region{ or in the subgiant region. To these,
  one merger resulting from the model 0.8675+0.4(2.3) can be added
  because it was formed by the Darwin instability just before 13 Gyr}. Its
surface layers may also show carbon and oxygen
depletion. Altogether, less than 10\% of the plotted BSs are expected to
show an underabundance of carbon and oxygen, which means that
the probability of finding
such an object among a random sample of BSs{ does not exceed 0.1}.

The observations of GCs agree with these
expectations. They are summarized by \citet{fer14} and
\citet{lov14}. C and O depletion for a few BSs in 47 Tuc, M30 and
$\omega$ Cen was detected, but none of the observed BSs in three other
clusters showed the expected alterations. The authors estimated that
only about 10\% of BSs may show CO depletion. That was fewer than they
expected, therefore they
speculated that the low percentage may result from the transient
character of the alterations. While this assumption may still be
viable, it is not needed to explain the observations.


\section{Discussion and conclusions}

{ We computed an extended set of CCBMs
with initial parameters expected to reproduce observations of CBs,
NCBs, and BSs in GCs. The initial component masses were between
0.7-0.9 $M_{\sun}$ for the primary and 0.4-0.8$M_{\sun}$ for
the secondary components. The initial orbital period distribution had a short-period
cut-off at 1.5 d resulting from the accepted binary formation
mechanism \citep{boss93,bonn94,krat10,mach08} and a maximum between
2-3 d resulting from the KCTF mechanism operating on suitably shaped
triples \citep{eggl06,per09}. The metal abundance was fixed
at Z=0.001. Magnetized winds from both components were adopted as the
driving force for orbit evolution. Evolutionary calculations were
carried out to the age of 13.5 Gyr, but detailed analysis of the
evolved objects was performed for three ages: 11, 12, and 13 Gyr.

The results show that all binaries with primary masses lower than
0.7$M_{\sun}$ and almost all with masses 0.75 $M_{\sun}$ remain
detached untill the age of 13.5 Gyr. On the other hand, almost all
binaries with initial primary masses close or equal to 0.9
$M_{\sun}$ complete the evolution as mergers or common-envelope products
before they reach the age of 11 Gyr. Only binaries with primary
masses embracing the TO mass at a specified age of 11, 12 or 13 Gyr
and with initial periods from a narrow interval around 2 d form CBs
or NCBs with periods shorter than 1 d, which should be observed as W
UMa-type stars in a GC of that age. Most of these variables do
not survive for more than 1 Gyr and either form a merger or an
Algol-type star with a period of several days. Consequently, only
4-5\% of the modeled binaries are in contact or near-contact
at each specified age. Observations show, indeed, that W UMa-type
binaries are rare among GC members \citep{ruc00}. Moreover, about 90\%
of model CBs lie close to, but below TO point on the HRD, which
means that on
average, only one in ten CBs falls in the BS region. Considering the
faint limit for CBs, the absence of model CBs fainter than $\sim$1 mag below
TO point agrees well with observations of GCs
\citep{kath09,kaetal10,kaetal13a}.

As the calculations show, freshly formed mergers have masses
between 0.9-1.5 $M_{\sun}$, as compared to a typical TO mass of $\sim
0.8 M_{\sun}$ in a GC. Thus, all they arrive in the BS region. In
effect, they outnumber binary BSs by at least an order of magnitude,
so among the random sample of BSs, fewer than 10\% are expected to be
close binaries. The observations support this result, although substantial
fluctuations of this fraction occur among individual clusters
\citep{fer09,dal13b,kaetal10,kaetal13a}.  Algols with periods longer
than 1 d occur predominantly to the red of the BS region because
of the
influence of a cool, Roche-lobe filling component that reddens the
binary.

{ All mergers with masses lower than 1.1 $M_{\sun}$ are still
  unevolved and lie along ZAMS, thus forming a blue BS
  sequence. On the other hand, all binary BSs lie in the red part of
  the BS region, thus forming a red sequence. A gap
  separates the two sequences. The existence of such a gap has been
  noted in { two} GCs, but it was linked to a highly nonuniform
  BS formation rate \citep{fer09,dal13b}. Further investigations
  should resolve this discrepancy.}

The uncertainties of the CCBMs mostly stem from the uncertainty of the
coefficients in Eqs.~\ref{massloss}-\ref{amloss} and the accuracy of
the stellar models. An {\em \textup{increase}} of the AML rate by 30\% results in a
decrease of the orbital period at RLOF by about 0.2-0.3 d. This has a
different influence on binaries with RLOF occurring on the MS and occurring
beyond the TAMS. In the former, RLOF takes place earlier, when the primary
is less evolved, hence the mass ratio past rapid mass exchange is
higher by about 0.05-0.1. Component merging and its following
evolution also occur earlier. In effect, CBs observed at the specified age
originate from binaries with initial periods longer by the same
amount. In the latter, the
shorter period at RLOF has a negligible effect on the process of mass
exchange because the fast radius increase of the subgiant component 
changes the time of RLOF by a mere few time
steps. The mass ratio past rapid mass exchange also changes
little, so only the orbital period is shorter, again, by
the same amount. A {\em \textup{decrease}} of the AML rate by 30\% has the opposite
effect to the same extent on the binary parameters. In short,
the uncertainty in the AML rate of 30\% translates into an uncertainty of
about 10\% in the binary and merger parameters.  

A decrease of mass-loss rate by a factor of two has an even weaker
effect. The total mass lost of a CB is between 6-7\% of the initial binary mass. Its cut in half
shortens stellar evolutionary time scales by about 10\% so RLOF occurs
somewhat earlier, but the final parameters of the considered models
agree within 4-5\% with those presented above. An increase of the mass-loss rate by a factor of two influences the resulting
parameters stronger, but, again, the change is lower than 10\%. Only
a substantially higher mass-loss rate (e.\thinspace g., by a factor of 5)
would qualitatively change the developed picture. Spherically
symmetric mass loss always causes a period increase. For each AML rate
there exists a maximum value of the mass-loss rate such that for higher
mass-loss rates orbit widens. That would substantially change the
binary evolution. There are no accurate measurements of mass-loss
rates for low-mass stars, therefore an extrapolation of data for solar-type stars was used. The mass-loss rates from Eq.~\ref{massloss} give
between 0.5-1.5$\times 10^{-11} M_{\sun}$ per year for a 0.8
$M_{\sun}$ star at the ZAMS and at the TAMS, respectively, and 2-7$\times 10^{-12}
M_{\sun}$ per year for a 0.5 $M_{\sun}$ star. Other authors suggested similar
rates for low-mass dwarfs, for example, \citet{cran11}, or even
somewhat lower \citep[$\sim\,10^{-12} M_{\sun}$ per year,][]{van97}. It
seems therefore that a substantially higher mass-loss rate is unlikely.

An assessment of the uncertainties connected with stellar evolutionary
models is more difficult. Even the models of single stars differ
significantly among different authors because of different input physics
\citep{verb95,nataf12}. In particular, the timescales of the consecutive
evolutionary phases and radii of the models may differ by several
percent. Of the same order is the uncertainty in stellar radii,
resulting from magnetic inflation observed in rapidly rotating cool
dwarfs \citep{tor10}. Fortunately, this uncertainty has only a minor
effect on the final results. The age of RLOF and the merging time change
by several time steps when we change the stellar radius by 10\%, and
the period boundary between CBs and NCBs (which,anyway, we treat
together) shifts only little.

These considerations about binary BSs only relate to close
binaries. Any BS can also be a member of a long-period
binary. In fact, our assumptions require that most of these
objects possess (or possessed in the early evolutionary phases) a
tertiary component that shortens the inner binary period by the KCTF
mechanism, as suggested by \citet{per09}. These authors also
discussed
the observational consequences of this assumption. They compared the
timescales for dynamical disruption of triples in cores and outer
regions of stellar clusters with the timescale for a formation of a
close binary. Both time scales are functions
of the outer binary period. It follows from the comparison that for
$P_{\rm{out}} \le 10^4$ d dynamical encounters are ineffective in the
outer parts of a GC. Thus a large part of BSs should  occur there
in triples (before merging) or in binaries. Only those formed in
initially soft triples might be disrupted. Altogether, the fraction of
binaries and triples among BS should be higher than the overall binary
fraction in the environment where they are observed, and most should be hard. The much higher frequency of the dynamical interactions in
GC cores make the prediction more difficult. The timescale for the
close binary production is short enough for $P_{\rm{out}} \le 10^4$ d,
so the KCTF mechanism can still operate in a GC core, but a high
interaction rate results in a disruption, a change of the configuration or a
merger of the inner binary. The outer binaries are even more
susceptible to disruption or component exchange. In effect, the
fraction of binary BSs in a GC core is expected to be lower and they
will have different parameter distribution from those in low-density
regions. If the triple system is not disrupted, a second-generation BS may be formed when a first-generation BS,
considered here, transfers matter to the original tertiary after
reaching the red, or asymptotic giant branch. In effect, a population of
BSs with white dwarf companions is expected. The observations of the
old open cluster NGC 188 revealed that a significant fraction
of BSs has such companions \citep{gos14}. Unfortunately, no such
observations for GCs are available.

For very few individual CBs in GCs, sufficiently accurate parameters
are known. Although we did not attempt at present to find progenitors
that fully match the observations, we can always find a model within the
CCBM set that closely reproduces the observed parameters. For example,
the model with initial masses 0.89+0.8 $M_{\sun}$ and the initial
orbital period of 1.9 d evolves after 11 Gyr into a binary with masses
1.26+0.28 $M_{\sun}$ and a period of 1.18 d, which matches the
variable V60 from M4 very well, for which the most accurate data have been
determined (see Table~\ref{v60} and Fig.~\ref{allbss}). Adding age and
metallicity as additional constraints should result in a unique
initial model for this binary. 

One of our main results shows that the binary-originated BS cover
most of the region on the HRD where the observed BSs occur. The least
massive mergers lie
close to ZAMS. This is due to the low masses of the merging components
with hardly depleted core hydrogen. The product of merging is then
unevolved and can remain in the BS region for 2-3 Gyr before moving to
the red giant region. They are indistinguishable in the HRD from the
collision-originated mergers of two low-mass stars. The higher the
component masses, the more advanced evolutionally the merger, hence
the blue boundary of the BS sequence turns more and more away from
ZAMS (see Figs.~\ref{gc11}-\ref{gc13}). Evolutionary advanced mergers
with core hydrogen depleted and binaries past RLOF populate the red
part of the BS region.  Only about 6-10\% of the mergers are expected to
rotate rapidly, and all probably lie close to the blue BS
boundary. Similarly, a low fraction of BSs may show chemical alterations
resulting from exposing layers where the CNO cycle operated. They
should occur predominantly among binaries with low mass ratios where a
sufficiently large amount of mass has been transferred between the
components. These predictions agree with observations of a
few GCs  \citep{ferrar06,l10,l12,lov13}.

To conclude, none of the observed parameters of an
individual single BSs can uniquely point to its origin: binary or
collisional. The presence of binaries among BSs obviously indicates that
at least a fraction of these objects has a binary origin. Can this be
true for most BSs? We noted profound differences among
individual GCs regarding the properties of BSs, therefore a single picture
that would be applicable for all clusters seems improbable. Evolutionary modeling of
a GC, including stellar and binary evolution, should provide credible predictions about the relative importance of different
production channels \citep{hyp13,chat13}. On the other hand,
a statistical analysis of GCs should help to verify these predictions
\citep{mil12,sil13}. It seems at present that models of GCs favor
formation mechanisms involving binaries. \citet{hyp13} modeled a
cluster with $10^5$ objects,  20\% of which are binaries. They
considered ten different ways of BS formation. The results showed that
single-star collisions produce a negligible number of BS, 13.6\% are
formed by collisions involving binaries (single-binary and
binary-binary), and the rest by binary mergers. They noted, however,
that these results may not apply to more massive and denser clusters.

Somewhat different results were obtained by \citet{chat13}, who showed
that in a dense environment (above $10^3 M_{\sun}/\rm{pc}^3$) most of
the BS are formed by collisions mediated by binaries, but in lower-density
clusters most BS (up to 60\%) result from mass transfer in
binaries. In effect, we can expect a varying proportion between both
kinds of BSs among different clusters. This is in line with
\citet{fer12}, who argued that BSs formed in cluster cores originate
from collisions, whereas those in cluster peripheries result from
binary evolution. Therefore a flat density distribution is seen in
dynamically young clusters, whereas an increasing concentration
toward the cluster center is visible in dynamically older clusters
when more massive BSs from the outer parts gradually migrate into the core. In
addition, \citet{li13} announced the observations of bimodal
distribution of BSs in the Large Magellanic Cloud GC Hodge 11. They
noted that BSs from the central distribution peak lie within the blue
sequence in the HRD, whereas those from the outer peak are placed in
the red part of the BS region. Assuming that the position in the HRD
uniquely determines an origin of a BS, they concluded that core BSs
are collisionally formed, whereas binary-formed BSs dominate in the
outer parts of the cluster. The results of the present paper show,
however, that this assumption is not correct. We also note that the
statistical comparison of the BS population in several GCs with the
simplified formation model clearly indicates that binary-formed BSs
dominate those collisionally formed  everywhere \citep{lei11b}.

In brief, the present evolutionary model of cool close binaries fully
explains the origin, evolutionary stage, statistical properties, and
individual physical parameters of CBs and NCBs observed in GCs.

Our results favor a  binary origin of BSs, but this is still an
open question. More observations of these stars in different regions
of GCs, in particular determinations of masses and multiplicity
fraction compared to the environment, will shed light on their origin.
Better evolutionary GC models will help to interpret observations and
to understand substantial differences observed among different GCs.}

\begin{acknowledgements}
We thank an anonymous referee, whose thorough report with many
relevant remarks and suggestions resulted in a substantial improvement
of the presentation of our results. We also thank Slavek Rucinski for
reading and commenting on the earlier version of the paper.
This research was partly supported by the National Science Centre
under the grant DEC-2011/03/B/ST9/03299.
\end{acknowledgements}

\bibliographystyle{aa}

\end{document}